\documentclass[10pt,journal,compsoc]{IEEEtran}
\ifCLASSOPTIONcompsoc
  \usepackage[nocompress]{cite}
\else
  \usepackage{cite}
\fi
\ifCLASSINFOpdf
\else
\fi

\usepackage{graphicx}
\usepackage{subfigure}

\usepackage{amsmath}

\usepackage{cite}
\usepackage{enumitem}

\usepackage{amsmath}
\usepackage{enumitem}
\usepackage{amssymb}

\usepackage{multirow}

\usepackage{program}
\usepackage[linesnumbered,lined,ruled,commentsnumbered]{algorithm2e}
\usepackage{color}

\definecolor{turquoise}{cmyk}{0.65,0,0.1,0.3}
\definecolor{purple}{rgb}{0.65,0,0.65}
\definecolor{dark_green}{rgb}{0, 0.5, 0}
\definecolor{orange}{rgb}{0.8, 0.6, 0.2}
\definecolor{red}{rgb}{0.8, 0.2, 0.2}
\definecolor{blue}{rgb}{0.2, 0.2, 0.8}

\begin{document}
%
\title{Building Anatomically Realistic Jaw Kinematics Model from Data}
%
%
%
%

\author{Wenwu~Yang,
        Nathan~Marshak,
        Daniel~S\'{y}kora,
        Srikumar~Ramalingam,
        and~Ladislav~Kavan
\IEEEcompsocitemizethanks{\IEEEcompsocthanksitem  W. Yang is with the School
of Computer and Information Engineering, Zhejiang Gongshang University, China, Hangzhou
310012. This work was done when he is a visiting scholar at the University of Utah. \protect\\
Email: wwyang@zjgsu.edu.cn
\IEEEcompsocthanksitem N. Nathan, S. Ramalingam and K. Ladislav are with the University of Utah.\protect\\
Email: \{srikumar.ramalingam, ladislav.kavan\}@gmail.com
\IEEEcompsocthanksitem D. S\'{y}kora is with the Czech Technical University in Prague, Faculty of Electrical Engineering. \protect\\
Email: sykorad@fel.cvut.cz}
\thanks{}}

%
%

\markboth{Building Anatomically Realistic Jaw Kinematics Model from Data, 11~May~2018}%
{Yang \MakeLowercase{\textit{et al.}}: Building Anatomically Realistic Jaw Kinematics Model from Data}
%



\IEEEtitleabstractindextext{%
\begin{abstract}
Recent work on anatomical face modeling focuses mainly on facial muscles and their activation which
generate facial expressions. In this paper, we consider a different aspect of anatomical face
modeling: kinematic modeling of the jaw, i.e., the Temporo-Mandibular Joint (TMJ). Previous work
often relies on simple models of jaw kinematics, even though the actual physiological behavior of
the TMJ is quite complex, allowing not only for mouth opening, but also for some amount of sideways
(lateral) and front-to-back (protrusion) motions. Fortuitously, the TMJ is the only joint whose
kinematics can be accurately measured with optical methods, because the bones of the lower and
upper jaw are rigidly connected to the lower and upper teeth. We construct a person-specific jaw
kinematic model by asking an actor to exercise the entire range of motion of the jaw while keeping
the lips open so that the teeth are at least partially visible. This performance is recorded with
three calibrated cameras. We obtain highly accurate 3D models of the teeth with a standard dental
scanner and use these models to reconstruct the rigid body trajectories of the teeth from the
videos (markerless tracking). The relative rigid transformations samples between the lower and
upper teeth are mapped to the Lie algebra of rigid body motions in order to linearize the
rotational motion. Our main contribution is to fit these samples with a three-dimensional nonlinear
model parameterizing the entire range of motion of the TMJ. We show that standard Principal
Component Analysis (PCA) fails to capture the nonlinear trajectories of the moving mandible.
However, we found these nonlinearities can be captured with a special modification of autoencoder
neural networks known as Nonlinear PCA. By mapping back to the Lie group of rigid transformations,
we obtain parameterization of the jaw kinematics which provides an intuitive interface allowing the
animators to explore realistic jaw motions in a user-friendly way.
\end{abstract}

\begin{IEEEkeywords}
Motion capture, motion processing, jaw kinematics, face animation.
\end{IEEEkeywords}}

\maketitle

\IEEEdisplaynontitleabstractindextext

%
\IEEEpeerreviewmaketitle

\IEEEraisesectionheading{\section{Introduction}}
Anatomical modeling of the face has been explored in the pioneering work of
\cite{terzopoulos1990physically,sifakis2005automatic}, but recent years witnessed a resurgence of
interest in anatomically-based facial animation
\cite{cong2015fully,lan2017lessons,kozlov2017enriching,Ichim2017}. Naturally, the primary focus is
accurate modeling of facial muscles and their ability to generate facial expressions. However, the
shape and expressions of the face are significantly affected by two major bones: the skull and the
mandible (lower jaw). This is evidenced by people who underwent jaw surgery, which is a relatively
frequent surgery to correct congenital malformations such as the overbite. The face after the
surgical treatment looks quite different and often much better than before the surgery. We argue
that realistic anatomically-based facial animation needs to start with an accurate jaw kinematics
model. With physics-based simulation of facial soft tissues, the relative rigid transformation
between the skull and the mandible has a significant effect on the result, because the bones are
used as Dirichlet boundary conditions. Even though, strictly speaking, the bones and their
attachment to the teeth is elastic, in normal physiological motions these deformations are
negligible and we can safely assume that hard tissues behave as rigid bodies. However, the rigid
motion of the mandible relative to the skull is not arbitrary, but is constrained by the anatomy of
the Temporo-Mandibular Joint (TMJ). The TMJ is a very complicated joint and enables functions such
as chewing of food or talking. Due to its complicated anatomy, the TMJ is also prone to pathologies
which are a common concern in medicine \cite{yoon2007kinematic}.

\begin{figure}[tb]
\centering
   \includegraphics[width=.8\linewidth]{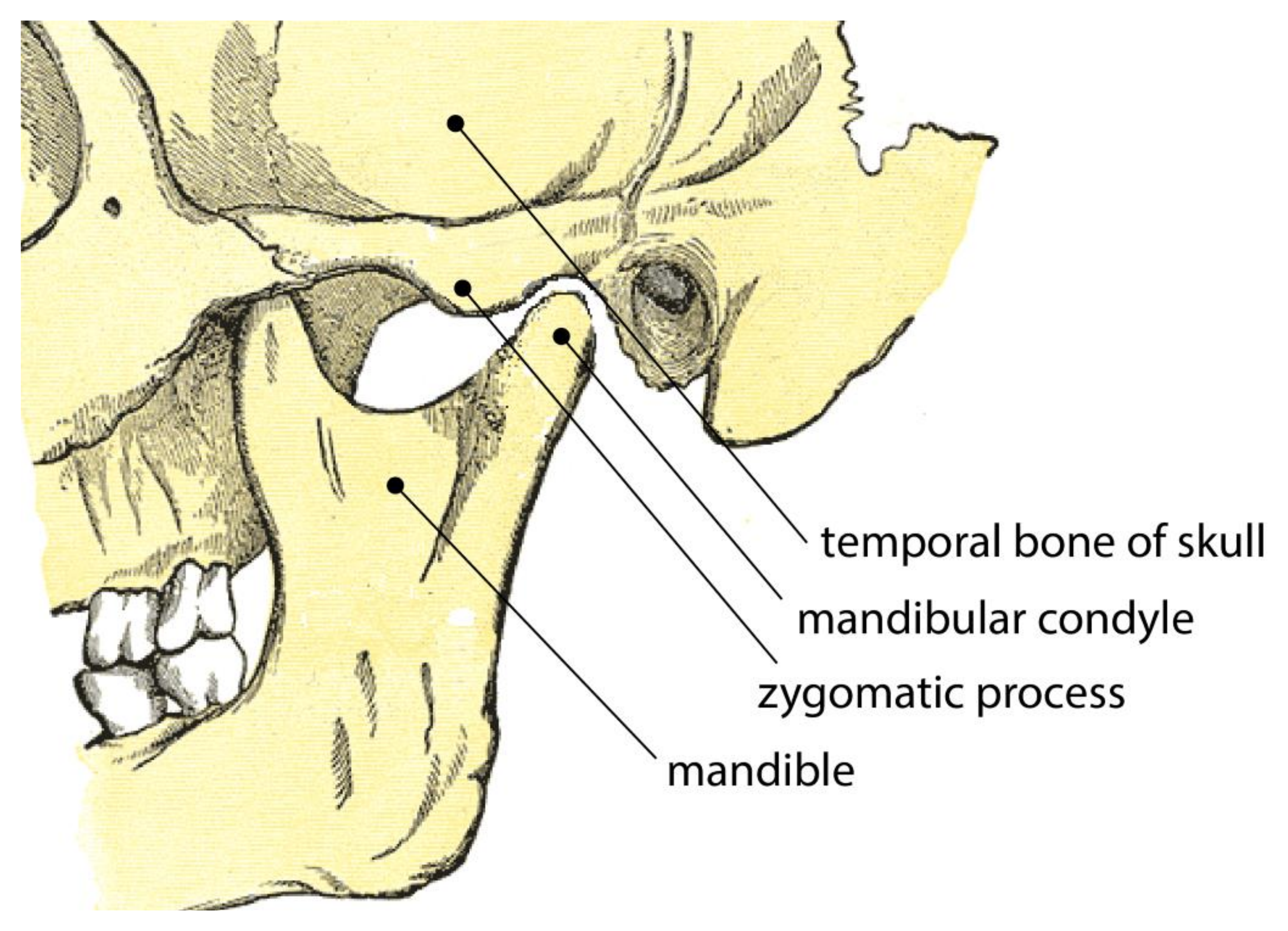}
   \caption{\label{fig:wiki_tmj}
     The temporalmandibular joint is the joint between the mandible and the temporal bone of the skull. }
\end{figure}

In this paper we focus on accurate modeling of the kinematics of the TMJ for the purposes of
computer animation. An ideal interface for animation (known as a ``rig'') should be user friendly.
The most prominent mode of motion is opening of the mouth. Even this common, everyday motion is,
kinematically, a non-trivial composition of rotation and translation (sliding). This sliding occurs
on a curve which reflects the geometry of the mandibular condyle and the zygomatic process which
are held in close proximity by connective tissues (Fig. \ref{fig:wiki_tmj} for anatomy of the TMJ).
Additionally, the jaw also allows for some amount of sideways and front-back translation, even
without opening the mouth. Normally, when the mouth is closed, the upper teeth rest naturally in
front of the lower teeth. We invite the reader to try translating their lower teeth forward -- they
can be moved \textit{in front of} the upper teeth. Similarly, it is also possible to move the lower
teeth from left to right. All of these motions are combined together to endow the jaw with its
basic functions, such as chewing or talking. Our goal is to provide an intuitive animation
interface allowing the users to explore realistic jaw motions. In particular, the users can
synthesize realistic jaw motions by just varying three parameters that correspond to anatomically
prominent modes of motion: opening, sideways sliding (known as ``lateral excursion'' in medical
terminology), and front-back translation (known as ``protrusion'' and ``retrusion''). In addition to synthesizing anatomically accurate jaw motions using an intuitive interface, our modeling will also allow us to validate if a given jaw motion is anatomically accurate or not.

\begin{figure}[tb]
\centering
   \includegraphics[width=.9\linewidth]{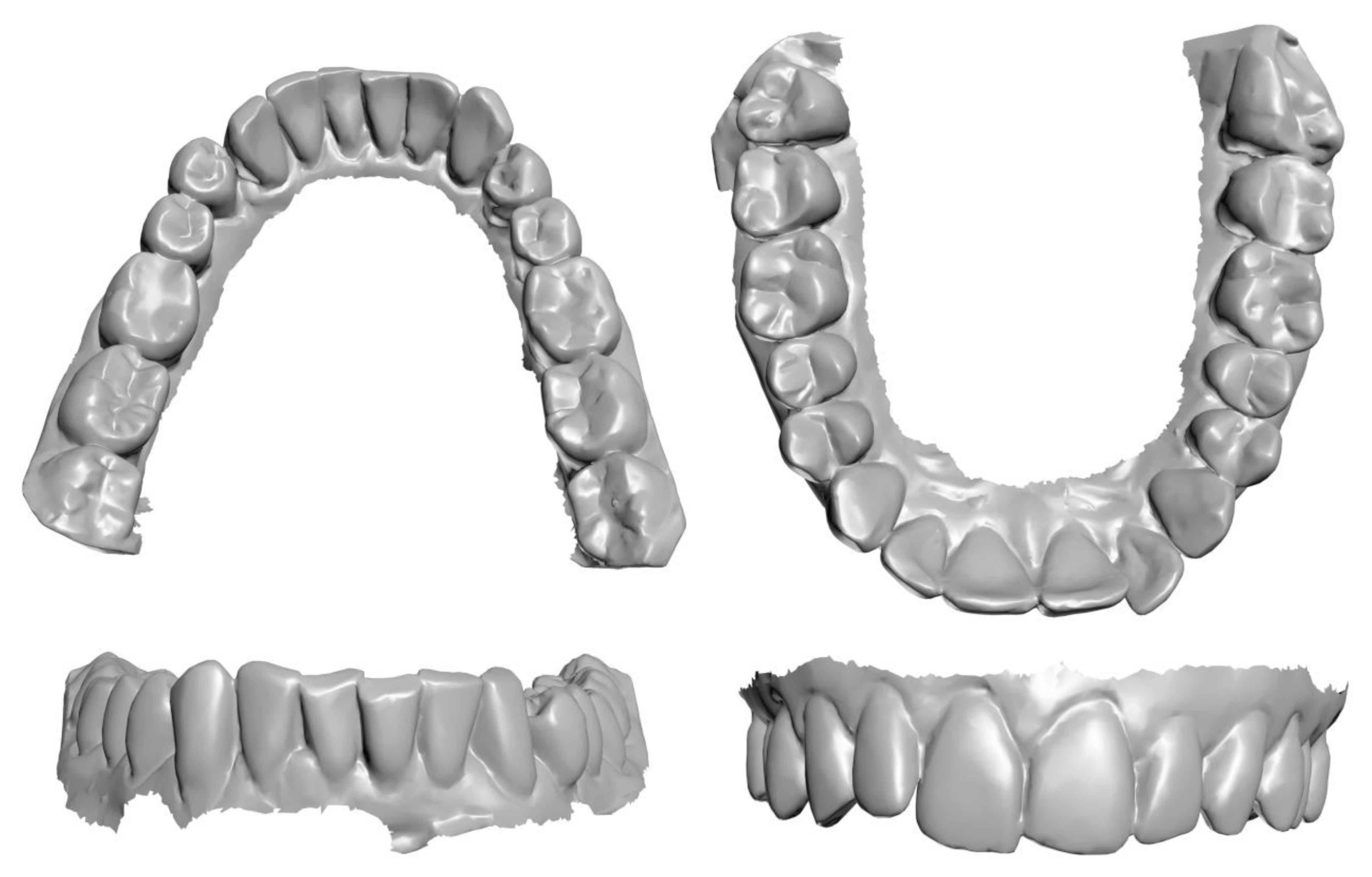}
   \caption{\label{fig:models}
     3D models (meshes) of the lower and upper teeth. }
\end{figure}

Previous kinematic models for the jaw were qualitative, designed by researchers who observed the
relevant anatomy and proposed models for its geometric behavior. We propose a different,
data-driven approach, taking advantage of the fortuitous fact that the kinematics of the jaw can be
measured with high accuracy using optical methods. This is because the skull and the mandible bones
are connected with the upper and lower teeth, and the motion of the teeth is directly visible if
the lips are open. To obtain data to train our model, we have asked an actor to exercise the entire
range of motion of the jaw while keeping the lips at least partially open. This performance is
recorded by multiple (we use three) cameras which have been calibrated (both intrinsically and
extrinsically, and also synchronized in time). We also create a highly accurate 3D models of the
teeth of the actor using a standard dental 3D scanner, see Fig. \ref{fig:models}. We use the 3D
models of the upper and lower teeth for tracking the video, reconstructing the 3D position and
orientation of the upper and lower teeth in each frame. Computing the relative rigid motion between
the upper and lower teeth poses in each frame gives us point samples of physiologically possible
jaw motions. These point samples are rigid transformations, i.e., points on the Lie group of rigid
motions, often denoted as $SE(3)$. The $SE(3)$ is a non-linear manifold due to the non-linearity of
rotations. To simplify our task of creating an intuitive parameterization of jaw kinematics, we map
our point samples from $SE(3)$ to the corresponding Lie algebra $se(3)$ via the logarithmic
mapping.
The Lie algebra $se(3)$ is a standard linear (vector) space and therefore permits standard
Principal Component Analysis (PCA). However, performing the PCA on our point samples in $se(3)$
fails to capture the nonlinearity of trajectories of motions such as mouth opening, where the
mandibular condyle slides along the zygomatic process along a curved path (see
Fig.~\ref{fig:PCAfail}).

In order to extract anatomically meaningful modes of motion from our input data ($se(3)$ samples),
we turned to autoencoder neural networks. In particular, we applied a special type of autoencoder
termed Nonlinear PCA (NLPCA) \cite{Scholz2008}. A modified version of NLPCA allowed us to explain
all of our input data with a generative model with only three parameters, corresponding to the main
modes of jaw motion: 1) mouth opening, 2) lateral excursion, 3) protrusion and retrusion.
Furthermore, we use our data to obtain explicit boundaries on each of these three modes of motion,
with the bounds on lateral excursion and pro/re-trusion depending on the amount of mouth opening.
The result is an intuitive and anatomically-realistic 3-parameterization for jaw kinematics.
There are two potential applications of our resulting jaw kinematics model: (1) a control
interface allowing animators to intuitively synthesize meaningful jaw motions, e.g., for
special effects animation \cite{Ichim2017}; (2) allow computer vision researchers to
automatically discard invalid jaw poses while tracking recorded facial performances \cite{wu2016anatomically}.

\section{Related Work}
The need for jaw kinematic modeling was identified in previous work in computer graphics. Sifakis
et al. \cite{sifakis2005automatic} created a jaw kinematics rig by designing sliding tracks of the
condyles identified from magnetic resonance images. Because MRI is not always available and may be
even medically contraindicated, \cite{wu2016anatomically} proposed a geometric rig offering two
rotational degrees of freedom along a common pivot point and one translational degree of freedom
along a fixed axis. This model was slightly generalized by \cite{Ichim2017} who proposed using
three translational degrees of freedom instead of just one, in addition to the two original
rotational degrees of freedom. Li et al.~\cite{li2017learning} use an articulated model including
the neck and eyeballs, with three rotational degrees of freedom for each of the joints, including
the jaw. Our approach does not require MRI but still produces highly accurate data-driven jaw
kinematics model for a given actor.

The mechanical function of the jaw has been studied in biomechanics, often using full six degrees
of freedom for the rigid body motion of the jaw relative to the skull \cite{koolstra2002dynamics},
even though reduced models were also considered, e.g., \cite{de2007validation} who proposed a
planar constraint along which the condyles can slide, resulting in four degrees of freedom.
A simulation platform can be used to create computational models using variables for modeling
gravity, external forces, and jaw muscle activity~\cite{HANNAM2010191}. These models were shown to
be capable of predicting jaw movements for mundane, but complex actions like
chewing~\cite{Hannam08adynamic} or post-reconstruction surgery~\cite{HANNAM2010191}.

An established tool to study bone kinematics is fluoroscopy (X-ray videos). Fluoroscopic studies of
the temporomandibular joint kinematics have been carried out on rabbits
\cite{henderson2014functional}, but the use of ionizing radiation (X-ray) for research purposes on
humans is not acceptable. Fortunately, the motion of the mandible relative to the skull can be
captured using optical methods if the lips are at least partially open. Tracking in videos is a
well studied computer vision problem~\cite{Lepetit2005}, and popular methods include template-based
tracking such as Lucas-Kanade~\cite{Lucas1981}, active appearance models~\cite{Cootes2001},
feature-based tracking~\cite{Lowe2004}, and edge or boundary-based
tracking~\cite{Klein2006,Ramalingam2010,Liu2012,Wang2014}. A key challenge in teeth tracking and
pose estimation problem is that it violates many common assumptions that are commonly satisfied in
a tracking framework. In particular, teeth are usually textureless and often partially occluded.
Tracking methods such as Lucas-Kanade~\cite{Lucas1981} and active appearance
models~\cite{Cootes2001} require the object to remain free from occlusion during the tracking
process and undergo only relatively small appearance changes with respect to the original template.
Unfortunately, in practice teeth are usually highly occluded and their appearance changes
considerably due to glossy nature of enamel. More robust keypoint-based methods such as
SIFT~\cite{Lowe2004} are also hardly applicable since teeth are usually smooth and self-similar
therefore it is difficult to find sufficient number of distinct feature points that can be tracked
consistently.

There are a few tracking algorithms that are customized for
teeth~\cite{Aichert2012,Wang2014,Wang2017}, and these tracking methods are primarily developed for
augmented reality applications, where the goal is to overlay an AR image on the patient for
providing additional assistance during dental surgery. In ~\cite{Wang2014}, a stereo camera is used
to capture and reconstruct the 3D contour of a patients teeth. This 3D contour is registered with
the 3D model obtained using CT scans using Iterative Closest point (ICP) algorithm. Using this
registration, we can have augmented reality (AR) overlay of 3D image on the patient during dental
surgery. Nevertheless, these techniques still assume avoidance of teeth occlusion.

In this work, we are primarily interested in principal component analysis (PCA) on 6-dimensional jaw motions in $se(3)$. Since linear PCA fails to capture the non-linear trajectories of motions, we resort to non-linear PCA (NLPCA) methods~\cite{Kramer1991}. Popular non-linear methods include principal curves~\cite{Hastie1989}, locally linear embeding (LLE)~\cite{Roweis2000} and Isomap~\cite{Tanenbaum2000}. In particular, we show that the input data can be explained using only three parameters corresponding to three main modes of jaw motions using a special type of autoencoder termed Nonlinear PCA (NLPCA) \cite{Scholz2008}.

\section{Method}
\subsection{Data Acquisition and Preparation}
\begin{figure}[tb]
\centering
   \includegraphics[width=.9\linewidth]{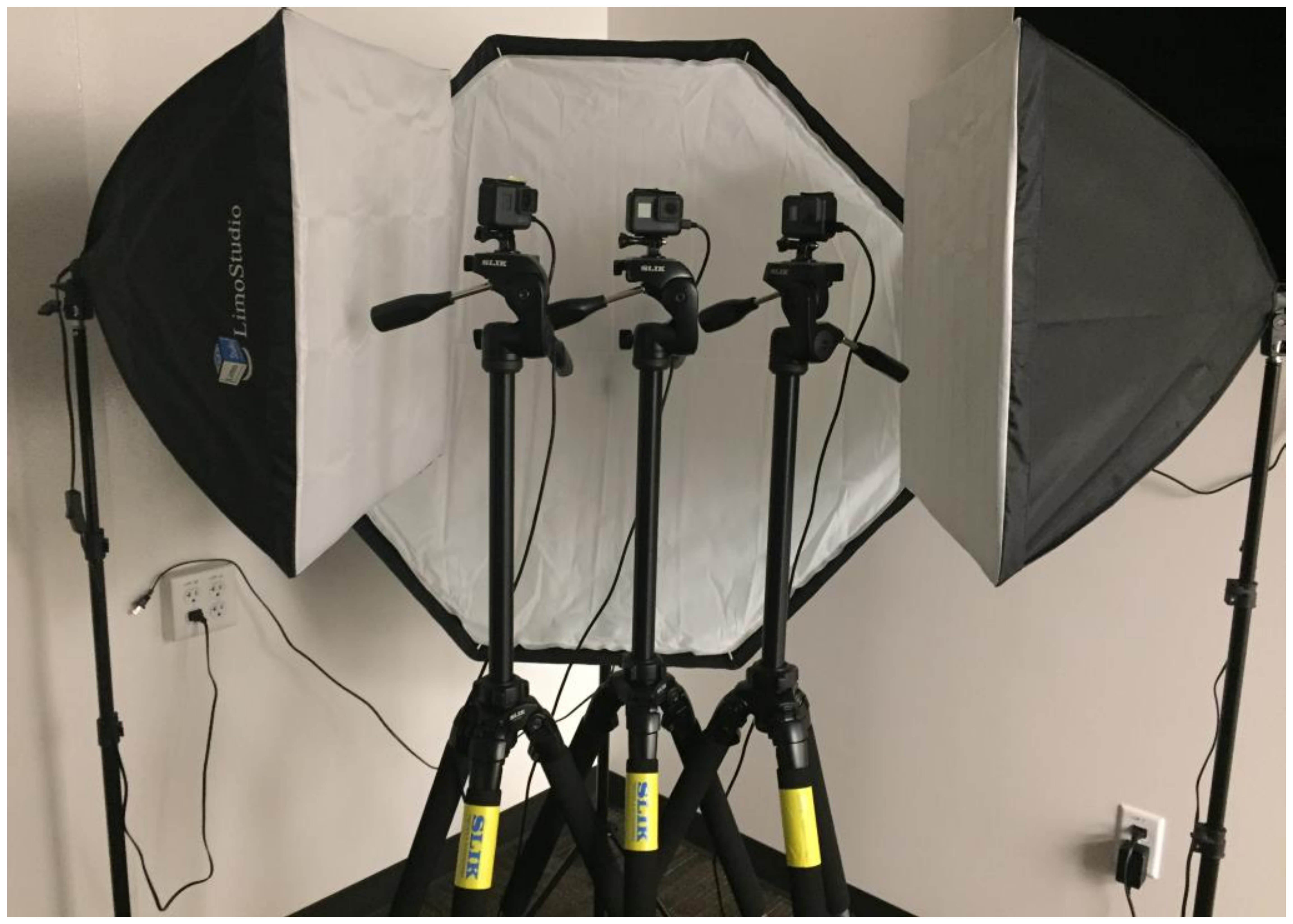}
   \caption{\label{fig:camera_rig}
     Our recording setup, featuring three tripod-mounted GoPro cameras and diffuse light sources. }
\end{figure}

\begin{figure}[tb]
\centering
   \includegraphics[width=.98\linewidth]{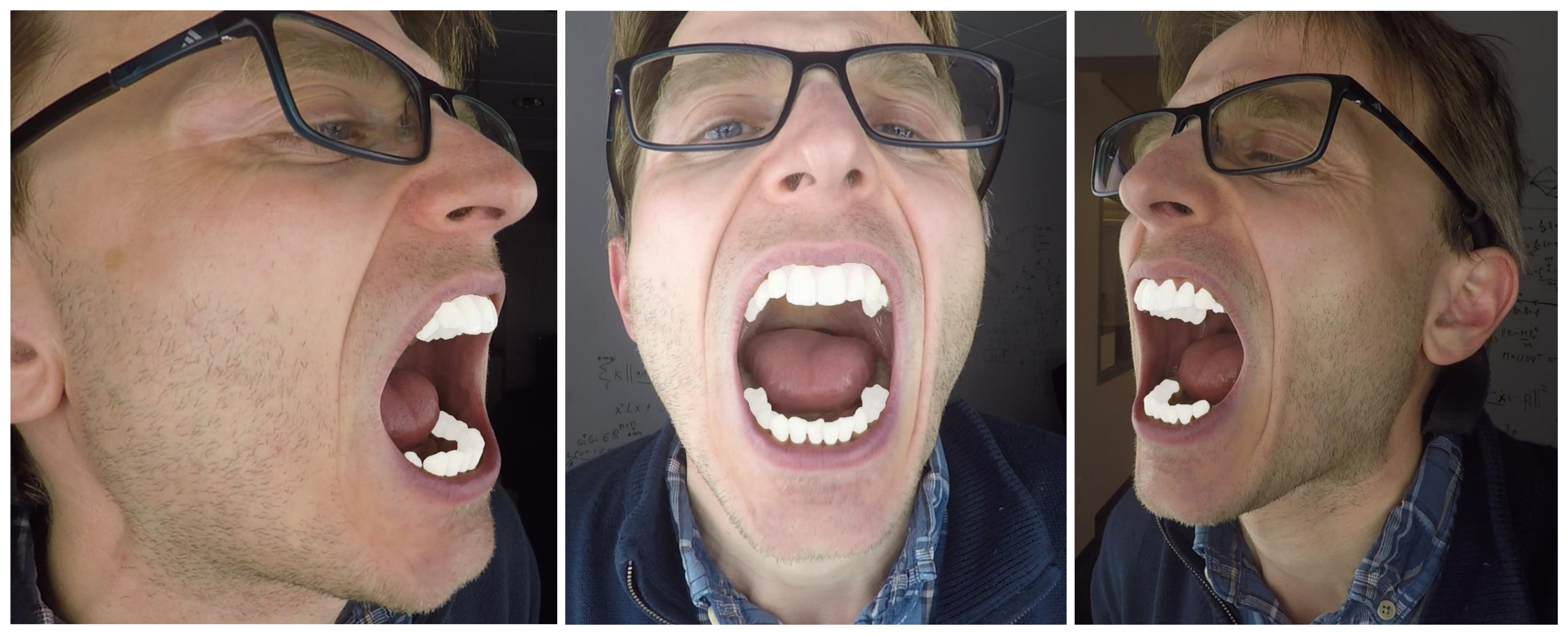}
   \caption{\label{fig:segmentation}
   Sample teeth segmentation results from synchronized images in three video cameras.}
\end{figure}

We start by obtaining the models of the upper and lower teeth of our actor by a standard dental
scanning procedure, producing detailed tooth geometry with distances in millimeters, see Fig.
\ref{fig:models}.
We capture the dynamic teeth performances using three tripod-mounted GoPro Hero 5 Black cameras,
see Fig. \ref{fig:camera_rig}. To reconstruct the teeth poses from the three camera videos,
we exploit the teeth's position and shape information that are implicitly encoded in the video
frames. We extract this information by segmenting out the teeth from the video frames, as
shown in Fig. ~\ref{fig:segmentation}. The segmentation of the video frames is performed by employing the Roto brush tool in Adobe Effect \cite{Bai2009}. The tool required minimal user interaction (i.e., only a few strokes) to achieve the teeth segmentation. Please note that gums between the teeth are allowed to be part of the teeth segmentation, since the color variation between them may be very small, as shown in Fig. \ref{fig:segmentation}. For additional technical details regarding our data capture and processing please see Section~\ref{sec:result}.

Only the relative motion between the skull and the mandible is relevant for further processing. The actual pose or orientation of the head is not important. To remove the global pose of the head from our study, we proceed as follows.
For a given image, let $(R^L, t^L)$ and $(R^U, t^U)$ be the rigid motion (rotation, translation) of the lower and upper teeth, respectively. Let us denote $(R, t)$ to represent the relative motion between the skull and mandible bones and it can be derived as follows:

\begin{equation}
\label{eqn:jaw_motion}
\begin{bmatrix}
    R^U       & t^U  \\
    0       & 1
\end{bmatrix}
\begin{bmatrix}
    R       & t  \\
    0       & 1
\end{bmatrix}
=
\begin{bmatrix}
    R^L       & t^L  \\
    0       & 1
\end{bmatrix}
.
\end{equation}
From Eqn. (\ref{eqn:jaw_motion}), we can compute
\begin{eqnarray}
R &= (R^U)^TR^L \\ t &= (R^U)^T(t^L-t^U)
\end{eqnarray}

\subsection{Learning Jaw Kinematics from Data}
Let $(R_i,t_i)$ be the relative jaw transformation of the frame indexed by $i$ from the input video. Our goal is to construct an anatomically realistic jaw kinematics model which can explain all of the measured relative jaw transformations $\{(R_i, t_i)\}_{i=1}^n$  (We used $n=833$ frames from our input training video) while providing the user with intuitive parameters for controlling the jaw's poses.

Each of our input data points $(R_i, t_i)$ lies on a $SE(3)$ manifold (the manifold of rigid
body transformations), which is nonlinear and thus not ideal for further analysis.
Therefore, we map each of our data points from $SE(3)$ to the corresponding Lie algebra $se(3)$
using the logarithmic mapping, which has a closed-form expression in the case of $SE(3)$
\cite{Murray1994}. Geometrically, this corresponds to unfolding the relevant part of the non-linear
manifold $SE(3)$ to a linear space (the Lie algebra). All of our subsequent analysis will be
performed on the data points in $se(3)$. We denote the resulting vectors as $\{v_i\}_{i=1}^m$,
where $v_i \in \mathbb{R}^6$.

Our first attempt to analyze the input training data $v_1, \dots, v_m$ is to apply Principal
Component Analysis (PCA), producing six principal components $\textbf{p}_{i}\in
\mathbb{R}^6$ with $i=\{1,2\ldots6\}$, where their corresponding six singular values are given by the set $\{10.38, 0.73, 0.44, 0.34, 0.25, 0.07\}$.
%
With this PCA basis, each jaw pose can be written as
\begin{equation}
\label{eqn:linearComb}
\textbf{p} = \sum_{j=1}^6 w_j \textbf{p}_{j},
\end{equation}
where $w_j \in \mathbb{R}$ are weights (PCA loadings) of each of the principal components. The resulting
$\textbf{p} \in se(3)$ can be mapped to $SE(3)$ by computing the exponential mapping
(closed-form expression \cite{Murray1994}). The resulting rigid body transformation can be then
used to transform the mandible mesh, producing visualization of the jaw motion represented by
the linear combination of the principal components.
\begin{figure}[tb]
    \centering
    \subfigure[]{
     \includegraphics[width=.95\linewidth]{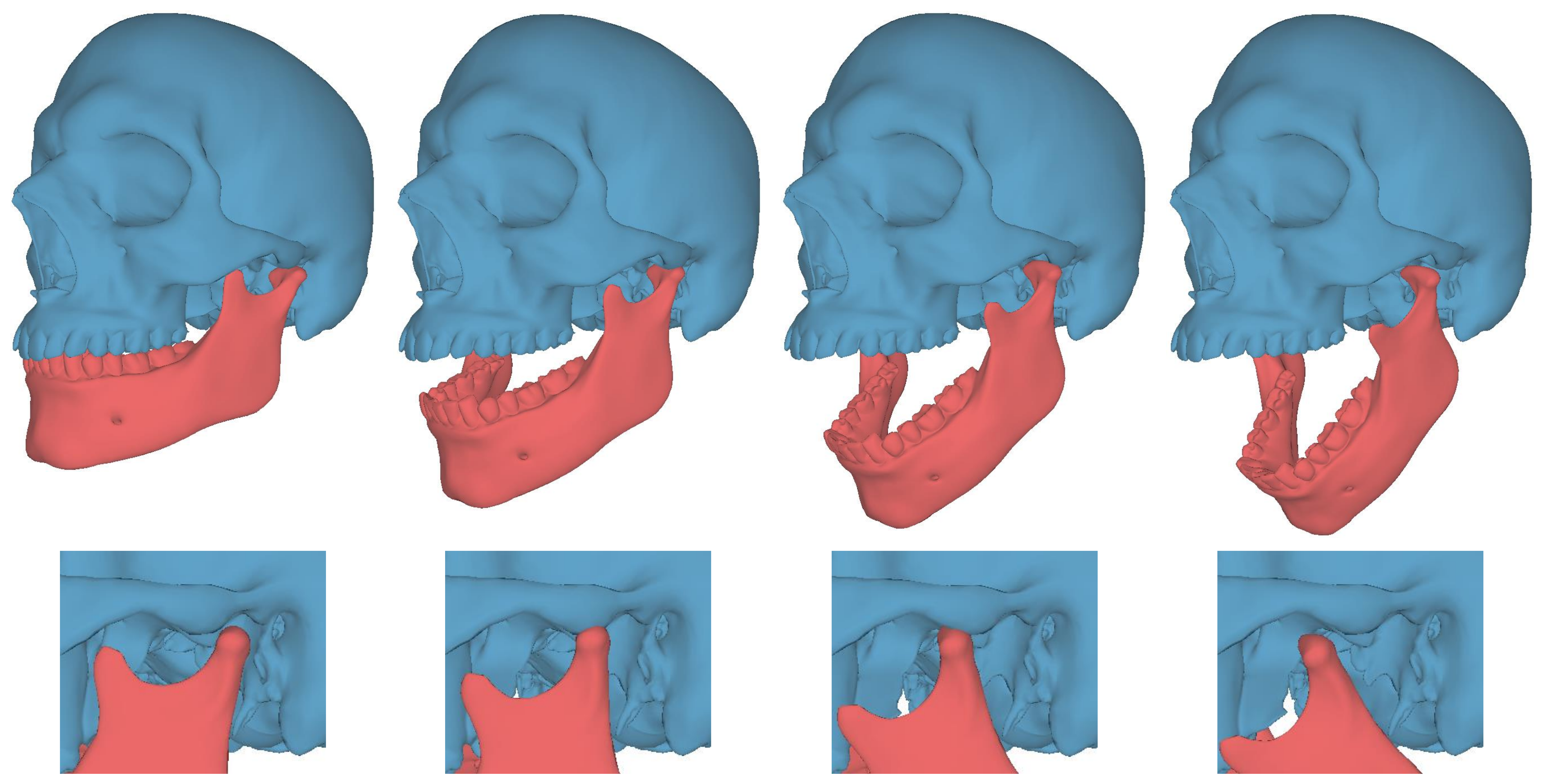}
    \label{fig:pca_linear}
    }
    \subfigure[]{
    \includegraphics[width=.95\linewidth]{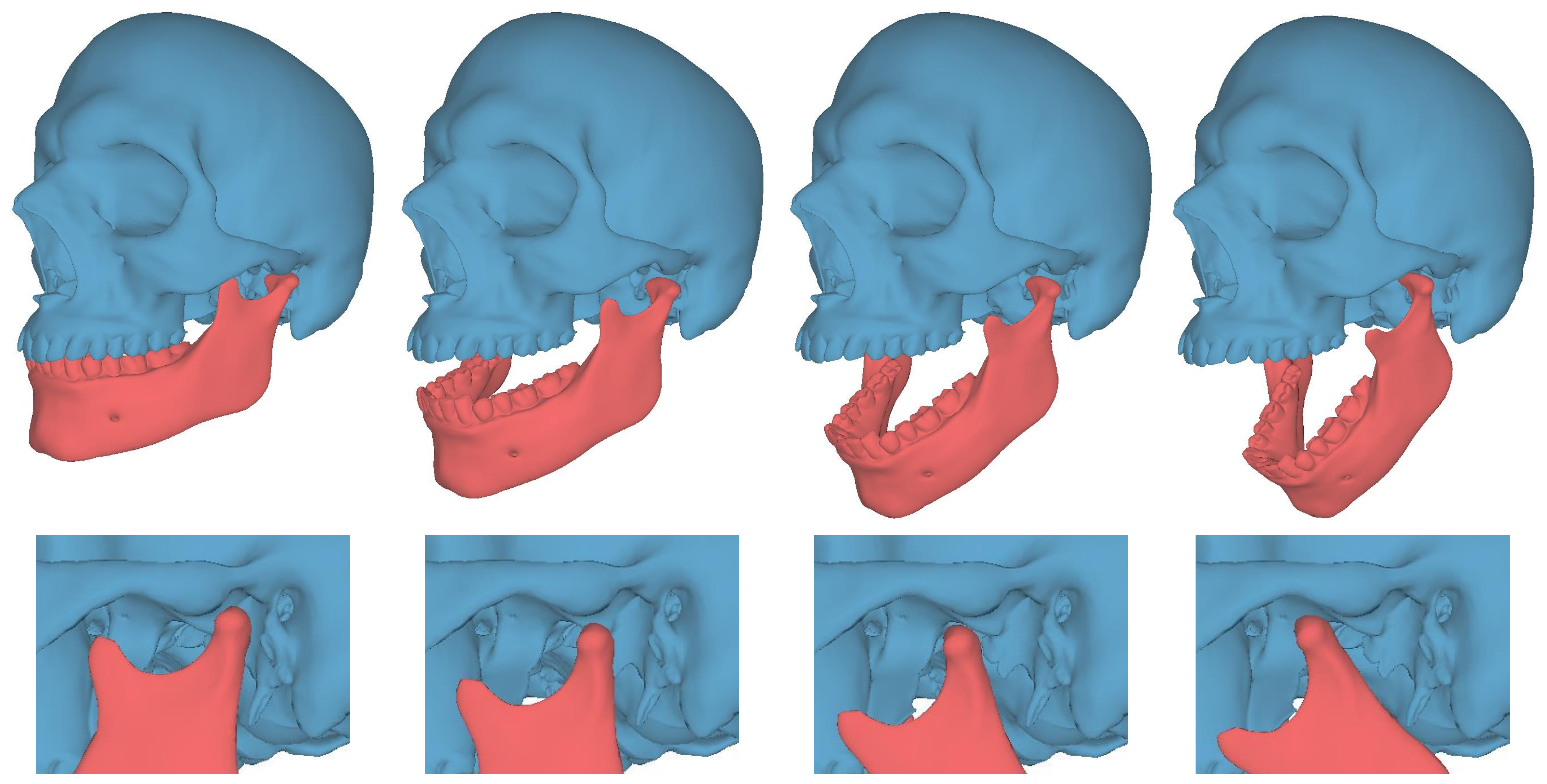}
    \label{fig:nonlinearpca}
    }
    \caption{Motions of mouth opening that are generated using PCA (a) and NLPCA autoencoder (b). The jaw poses in (a) are generated by scaling the first principal component: $w_1\textbf{p}_{1}$, for $w_1 = \{-1.6157, -0.5717, 0.4723, 1.5165\}$ (left to right), while the jaw poses in (b) are generated as $\Phi_{gen}(c_1, 0, 0, 0, 0)$ where $c_1$ is $\{-0.1943, -0.0195, 0.1552, 0.3300\}$ (left to right). Note that the
     zoom out in the bottom row of (a) illustrates that the PCA does not learn the correct non-linear
     trajectory and the mandibular condyle intersects the zygomatic process, while in (b) the condyle slides along the zygomatic process, which is the desired anatomically-realistic behavior. }
    \label{fig:PCAfail}
\end{figure}

%
In Fig. \ref{fig:pca_linear}, we visualize the effect of the first principal component
$\textbf{p}_{1}$ which, not surprisingly, captures the dominant motion -- opening of the
mouth. However, the visualization in Fig. \ref{fig:pca_linear} also reveals the limitations of PCA (please see also the animation in the accompanying video). Specifically, even though the jaw motion generated by scaling the first principal component roughly corresponds to opening of the mouth and includes both rotation and translation, it fails to accurately capture the trajectory of the mandibular condyle. When opening the mouth, the mandibular condyle slides over the surface of the zygomatic process (see Fig. \ref{fig:wiki_tmj}). Because the zygomatic process is curved (as opposed to flat), this results in a non-linear trajectory. The non-linearity of this trajectory can be observed by real-time magnetic resonance imaging of a human subject performing voluntary mouth opening and closing \cite{Zhang2011} (please see the accompanying video). Unfortunately, the first principal component produces only a crude approximation of this nonlinear trajectory (specifically, this approximation corresponds to a straight line in $se(3)$). There are other limitations with PCA. The second and the remaining principal components do not correspond to anatomically meaningful motions such as lateral  excursions  and  pro/re-trusion. It is also challenging to generate anatomically realistic motions by considering the 5 or 6 principal components within their boundary values.

To be able to capture the correct anatomical behavior of the temporomandibular joint during mouth opening with a single parameter, we turn to a more powerful, non-linear data analysis tool: autoencoder neural networks \cite{Goodfellow2016}. We
trained various autoencoder architectures using TensorFlow \cite{Abadi2016} and found that
substantial dimensionality reduction can be achieved, e.g., using only three dimensions can
reproduce the training data $\{v_1, \dots, v_m\}$ very accurately (unlike PCA, which produces
relatively large error with only three principal components). Even though autoencoders
without any non-linear units are almost equivalent to PCA \cite{Baldi1989}, there is a
catch in the ``almost.'' Specifically, the reduced parameters produced by the encoder part of an
autoencoder do not have any hierarchical meaning as is the case in PCA. In PCA, the first principal
component explains the largest variance in the data, but this is not the case for the first
component produced by the encoder. Fortunately, we found that a solution to this problem has
already been described by Scholz and colleagues \cite{Scholz2002,Scholz2008}, who proposed a
modified autoencoder network which mimics the hierarchical property of the principal components.
Their approach is called NLPCA (for Non-Linear PCA).



Let $\mathcal{S}$ be a data space given by our 6-dimensional data points in $se(3)$ and
$\mathcal{P} \subseteq \mathbb{R}^n$ a component space with $n\leq 6$. NLPCA learns two nonlinear
functions $\Phi_{extr}$ and $\Phi_{gen}$, where $\Phi_{extr}: \mathcal{S} \mapsto \mathcal{P}$ is
called \emph{component (or feature) extraction} function (corresponding to the encoder part of an
autoencoder) and $\Phi_{gen}: \mathcal{P} \mapsto \mathcal{S}$ is called \emph{data generation}
function (corresponding to the decoder part). The extraction function $\Phi_{extr}$ transforms a
6-dimensional data point into the corresponding component representation
$\textbf{c}=(c_1,c_2,\ldots,c_n)$, while the generation function $\Phi_{gen}$ performs the reverse,
i.e., reconstructs the original data points from its lower-dimensional component representation.
With our training data $v_1, \dots, v_m$, we found that only five dimensions are needed with NLPCA;
the sixth dimension introduces only minimal modifications which can be attributed to noise in the
input data. Furthermore, visualizing the effect of the individual components, we observe
%
%
that the nonlinear characteristics of the mouth opening motion are effectively captured by first
component ($c_1$). Visualizing the results of $\Phi_{gen}(c_1, 0, 0, 0, 0)$ for varying $c_1$
produces anatomically realistic nonlinear trajectory of the mandibular condyles, see Fig.
\ref{fig:nonlinearpca}. Most importantly, the mandibular condyle now slides along the zygomatic
process, as opposed the unrealistic intersection produced by PCA (Fig. \ref{fig:pca_linear}). Even
though this five-dimensional model can accurately represent all possible motions of the jaw, we
found it is possible to reduce the number of parameters further, providing a more compact and
intuitive interface to the user while increasing the error only negligibly.

By visualizing the effect of the components $c_2$ and $c_3$, we found that both of them correspond
to lateral excursions (moving the jaw sideways, left and right). Similarly, the last two components
$c_4$ and $c_5$ correspond to forward and backward motion of the jaw, known as protrusion and
retrusion. This observation motivates the final refinement of our model, which we describe next.
%
%
%
\begin{figure}[tb]
\centering
   \includegraphics[width=.95\linewidth]{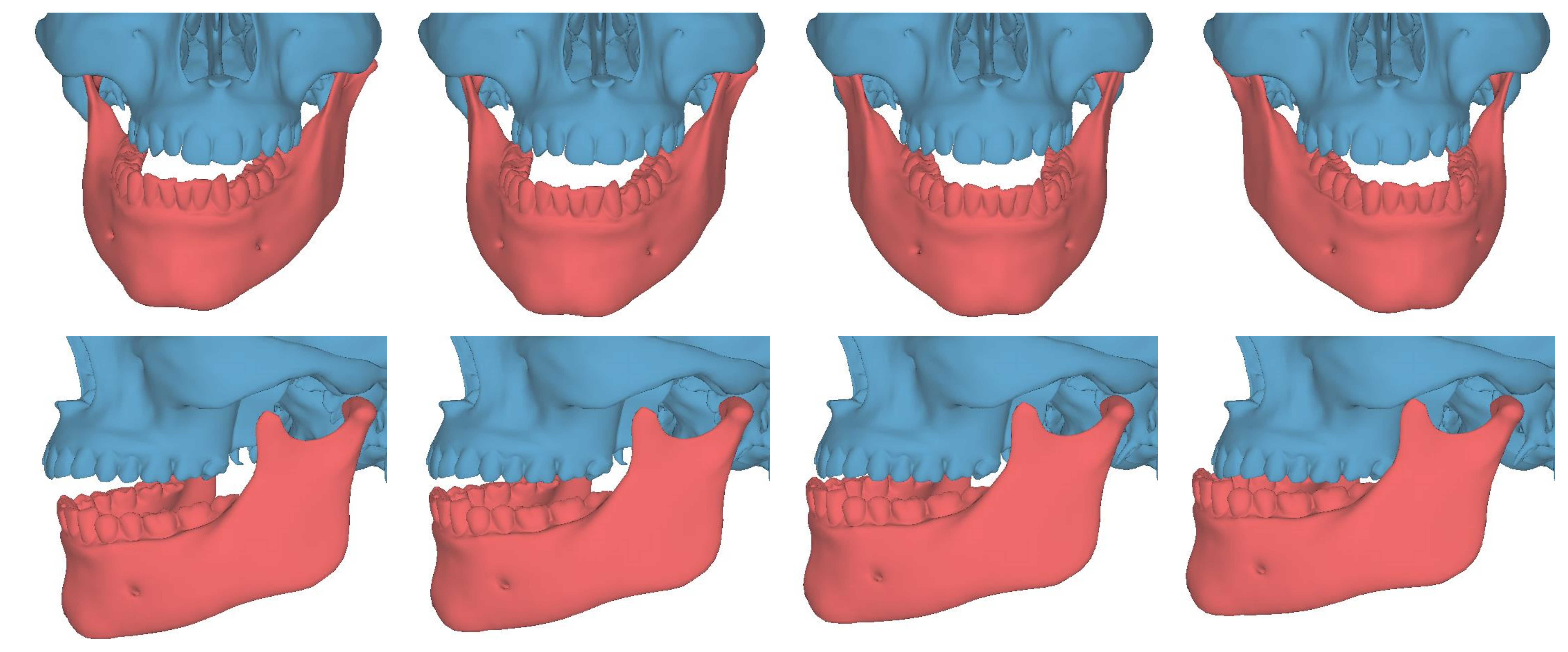}
   \caption{\label{fig:nonlinearpca_subData}
     Top row: lateral excursions learned by non-linearly reducing $D_{23}$ to a single dimension;
     Bottom row: the same approach applied to $D_{45}$ produces one-dimensional parameterization of
     protrusion and retrusion.}
\end{figure}
%
%
First, we generate a set of data points (denoted as $D_{23}$) which correspond to zeroing all
components except for $c_2$ and $c_3$, i.e., $\Phi_{gen}(0, c_2, c_3, 0, 0)$, where the values of
$c_2$ and $c_3$ are given by encoding of our training data.
%
Similarly, we construct another set of data points (denoted as $D_{45}$) in terms of the components
of $(0, 0, 0, c_4, c_5)$. Next, we run NLPCA \cite{Scholz2008} on the $D_{23}$ dataset to
nonlinearly reduce its dimensionality to one. We denote the resulting one-dimensional generation
function (decoder) as $\Phi_2(d_2)$, where $d_2$ is a scalar input parameter. We observe that by
varying $d_2$, the function $\Phi_2$ can explain the entire range of lateral excursions (sideways
motions of the jaw from left to right), during which the trajectory of the condyles again succeeds
to avoid inter-penetrations with the zygomatic process (similarly as before for jaw opening). Even
better, it turns out the similar trick works also for $D_{45}$! Again, we execute NLPCA on the
$D_{45}$ dataset and obtain a one-dimensional generation function $\Phi_3(d_3)$, such that varying
the scalar parameter $d_3$ produces an entire range of anatomically realistic protrusion and
retrusion. These motions are visualized in Fig.~\ref{fig:nonlinearpca_subData}. To make our
notation succient, let us introduce a shorthand $\Phi_1(d_1) := \Phi_{gen}(d_1, 0, 0, 0, 0)$ where
$d_1$ is a parameter corresponding to mouth opening.

Our final model parameterizing the jaw kinematics with only three dimensions is a linear
combination of the individual parts:
\begin{equation}
\label{eqn:synthesize}
\Phi(d_1, d_2, d_3) = \Phi_1(d_1) + \Phi_2(d_2) + \Phi_3(d_3)
\end{equation}
Linear combination is justified by the fact that the Lie algebra $se(3)$ is a linear space and the
maximal relative rotations are bounded (far below 180 degrees) \cite{Alexa2002}. As we discuss in more details in Section \ref{sec:result}, this final $\Phi(d_1, d_2, d_3)$ introduces only marginally higher error compared to the
five-dimensional NLPCA model, while using fewer parameters and being much more intuitive to the
user.

\begin{figure}[tb]
\centering
   \includegraphics[width=.95\linewidth]{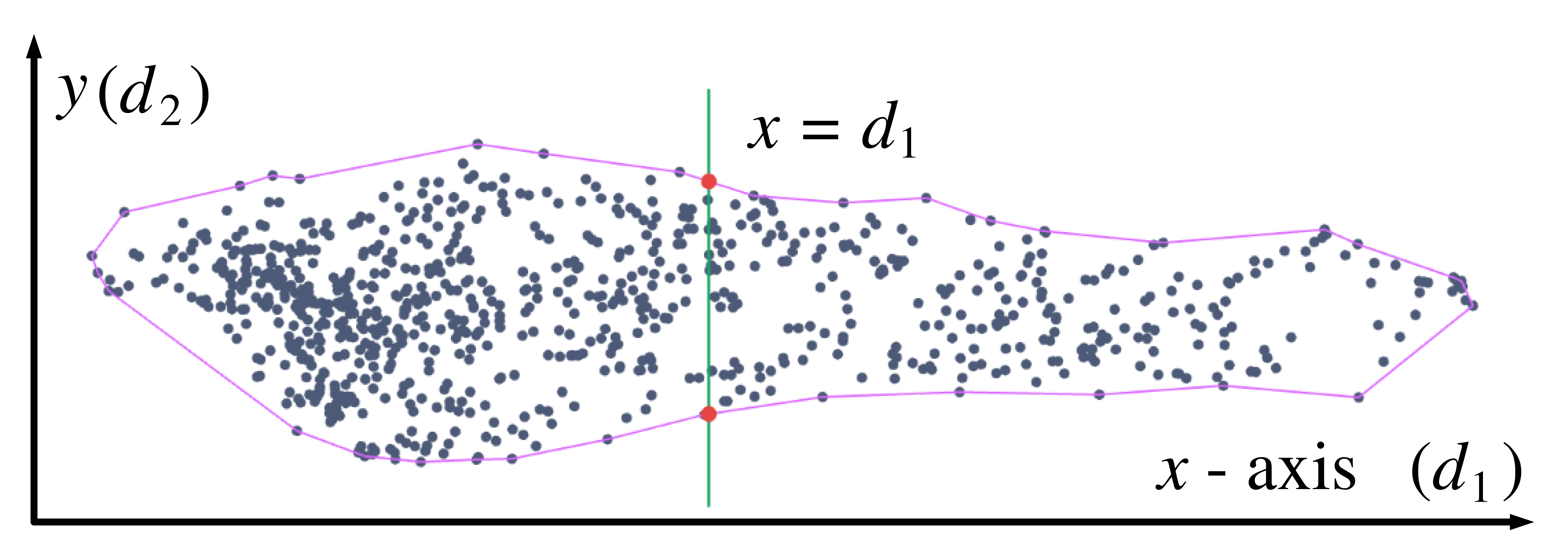}
   \caption{\label{fig:pca_nonlinear_boundary}
     The range of the parameter $d_2$ is determined with respect to a given $d_1$ (mouth opening).The parameters $(d_1, d_2)$ of our original data points are projected onto
     the $x$-$y$ plane and the valid range of $d_2$ is determined by $y$-coordinates of the
     intersection points between the line ($x=d_1$) and a polygonal boundary of the 2D projection points.}
\end{figure}

\emph{Valid Component Representation. }\quad It is evident that for a valid parameter
representation $\textbf{d}=(d_1,d_2,d_3)$, each of its parameters should be in a specified
range. Since each of our original data points has a corresponding parameter representation, a
simple way to determine the parameter ranges would be to bound each parameter using the
parameter representations of the given data points. However, in real jaw motion, the ranges of
parameters $d_2$ and $d_3$ are dependent on the current value of $d_1$. To
demonstrate this, we invite the reader to try sliding their lower teeth to the right and then
opening the mouth -- the lower teeth are automatically moved back to the center. Therefore, we need
to dynamically determine the valid ranges for the parameters $d_2$ and $d_3$, with respect to
current $d_1$. As shown in Fig. \ref{fig:pca_nonlinear_boundary}, we use $d_2$ as an example to
describe our solution. First, we project the parameters $(d_1, d_2)$ of our original
data points onto a $x$-$y$ plane. Then, a polygonal boundary is formed using the alpha shape ($alpha=0.12$ in our experiments) of these planar points. For a given $d_1$, the current range of $d_2$ is
determined by the intersection points between the polygonal boundary and the line $x=d_1$.


\section{Model-Based Teeth Motion Tracking}
\label{sec:tracking}
Essentially, our teeth motion tracking problem is a 3D-2D matching problem \cite{Gold97,Rosenhahn05} where the task is to find the best pose of a 3D model so that its 2D projections are well aligned with the 2D images. In our scenario,
the 3D model is a teeth model, corresponding to either the upper or lower teeth of the performer, and the model pose corresponds to a rigid motion. More formally, we denote the teeth model as $\mathcal{M}$ and represent the rigid motion as a $6$-tuple $T=\{\theta_x, \theta_y, \theta_z, t_x, t_y, t_z\}$, where $\theta_x$, $\theta_y$, $\theta_z$ are the rotation angles around the $x$-, $y$-, and $z$-axes, respectively, and $t_x$, $t_y$, $t_z$ are the translation offsets along the $x$, $y$, and $z$ directions, respectively. To make the projections of $\mathcal{M}$ most consistent with the (three in our case) images that corresponds to a common video frame, we can solve the following optimization problem:
\begin{equation}
\mathop{\arg\min}_{T}\quad \sum_{i=1}^3 E(\mathbb{P}_i(R\mathcal{M}+t), I_i),
\end{equation}
where $R=R_\mathbf{z}(\theta_z)R_\mathbf{y}(\theta_y)R_\mathbf{x}(\theta_x)$ with $R_\mathbf{x}(\cdot)$, $R_\mathbf{y}(\cdot)$, and $R_\mathbf{z}(\cdot)$ being the rotation matrices about the $x$-, $y$-, and $z$-axes, respectively, with a specified angle, $t=(t_x, t_y, t_z)$ is a translation vector, $\mathbb{P}_i(\cdot)$ is an operation that projects a 3D model to the image plane of the $i$th camera, $I_i$ corresponds to a video frame image that is captured by the $i$th camera, and $E(\cdot,\cdot)$ defines an energy function which measures the inconsistency or discrepancy between the 2D projection of the 3D model and the real 2D image. In our implementation, we use hardware pipeline via OpenGL to conduct the projection operation $\mathbb{P}_i(\cdot)$, with respect to the cameras' intrinsic and extrinsic parameters.





\begin{figure}[tb]
\centering
   \includegraphics[width=.8\linewidth]{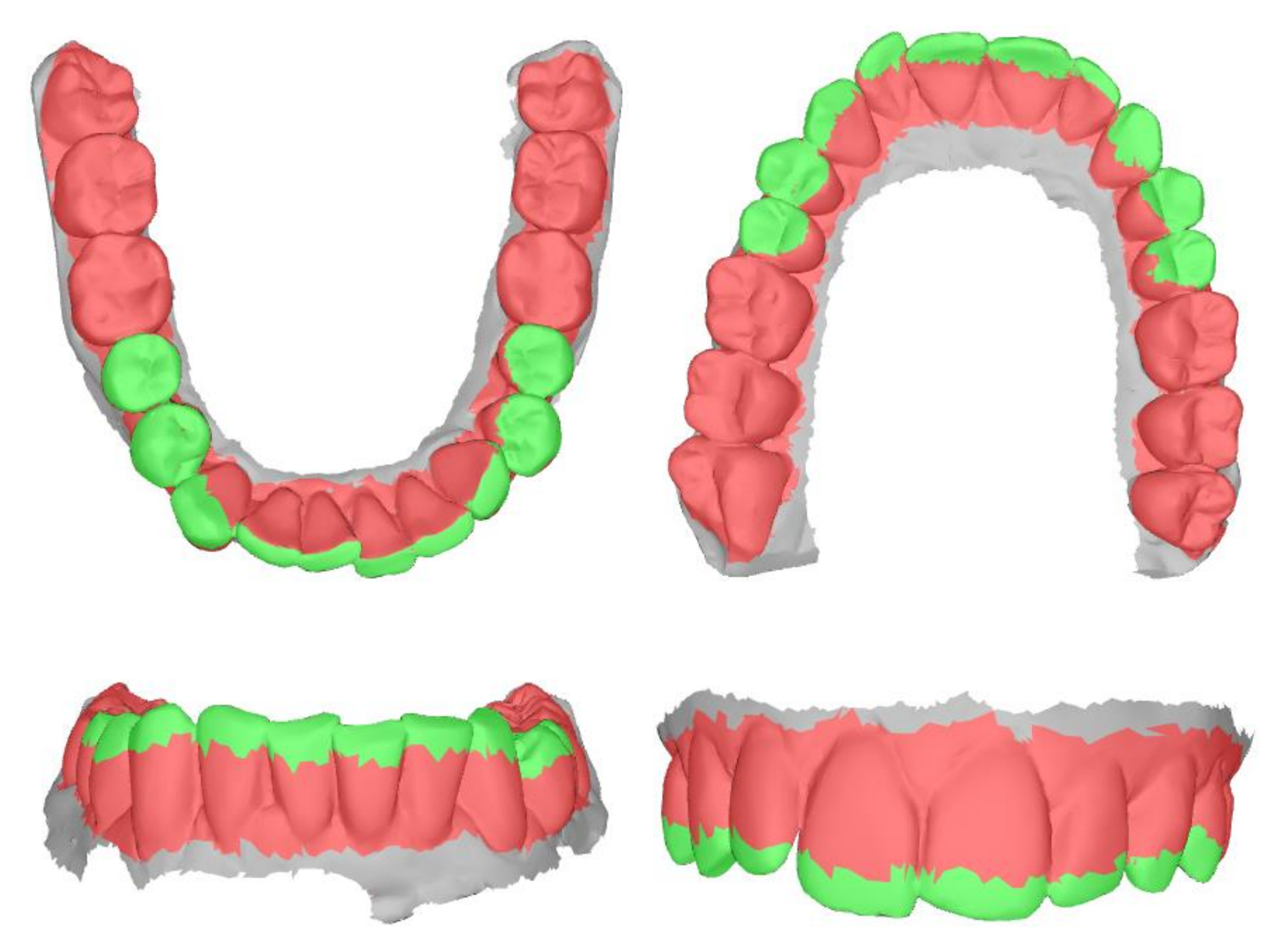}
   \caption{\label{fig:teethmark}
    The teeth model is roughly marked out from the 3D scans of the performer's  teeth rows. The geometry of the teeth model consists of the triangles in the red and green regions, where the green parts correspond to the boundary of interest of the teeth model. } 
\end{figure}

To define $E$, our idea is to fully exploit the geometry information of the teeth model (see Fig. \ref{fig:teethmark})  as well as the shape and position information of the performer's teeth that are encoded in the teeth segment of the captured images.  We define $E$ as follows:
\begin{equation}
\label{eqn:energy}
E = E_{SP} + \lambda E_{BND}.
\end{equation}
where $E_{SP}$ measures the overlap of the teeth model with the actual visible teeth segment while $E_{BND}$ ensures alignment of the roof region of frontal teeth which we call boundary of interest (BOI). We use fixed weight $\lambda = 0.4$ for all of the results in this paper.


\begin{figure*}[tb]
\centering
   \includegraphics[width=.95\linewidth]{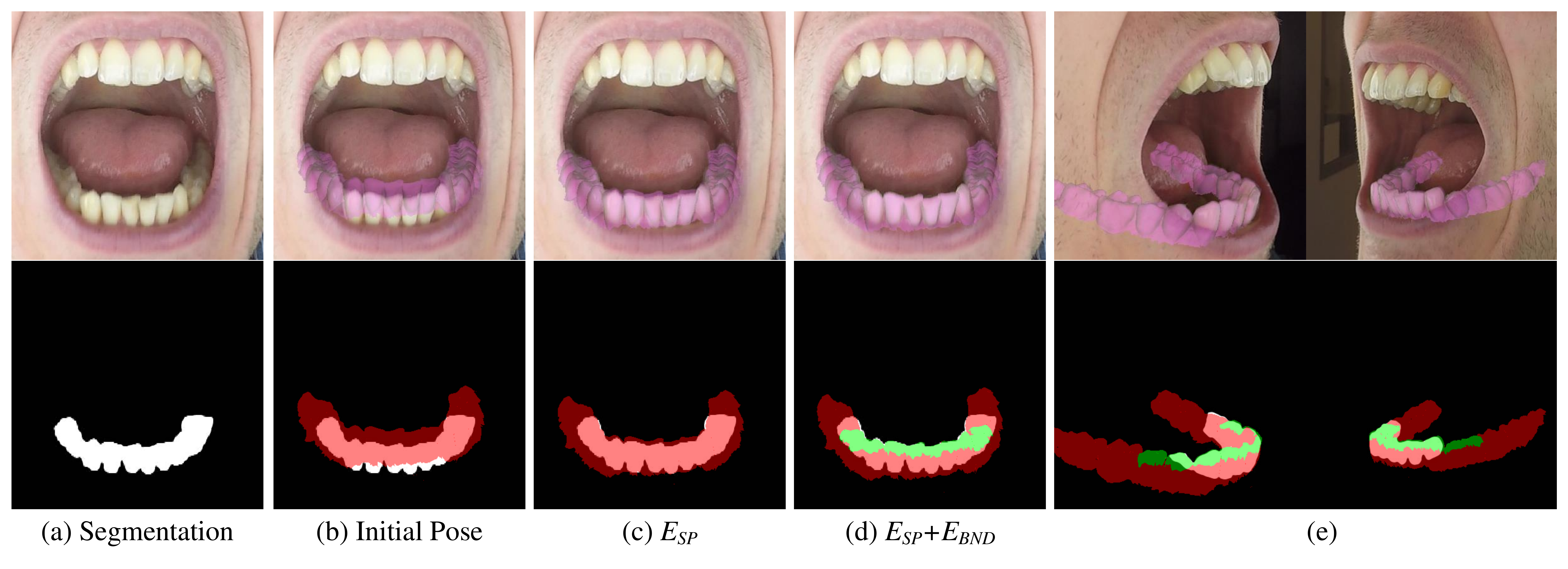}
   \caption{\label{fig:energy}
     Optimization energy of the (lower or upper) teeth pose. Given the captured images in Fig. \ref{fig:segmentation} and the initial pose of the teeth model, we render for each camera the 2D projection of the model's geometry and the teeth segmentation into the same screen buffer (b). The overlapping pixels are used to define energy terms (c,d) which measure how consistent is between the poses of the teeth model and the real teeth at the current video frame. In (d,e), the projections of the optimal teeth pose in all of the three cameras are shown.}
\end{figure*}

To compute $E_{SP}$ for every captured image we render its teeth segmentation with white color into clean (black) screen buffer (see Fig.~\ref{fig:energy}a). Then, the teeth model at current pose is rendered in red color (with alpha blending $\alpha=0.5$), i.e., projected with respect to the intrinsic and extrinsic parameters of the corresponding camera. Thanks to this second rendering pass the pixels of the teeth segmentation that are overlapped with the projection of the teeth model will change to pink color (see Fig.~\ref{fig:energy}(b)). The closer the poses of the teeth model are to the actual performer's teeth, the fewer pixels with white color remain. Therefore, we define the energy term $E_{SP}$ as follows:
\begin{equation}
\label{eqn:energy_sp}
E_{SP} = \#Pixel\_White,
\end{equation}
where $\#Pixel\_White$ refers to the number of white pixels in the screen buffer.


Due to the occlusion by lip or tongue, the teeth segmentation of the performer is usually smaller than the 2D projection of its teeth model, such that the energy term $E_{SP}$ may lead to sub-optimal result, as shown in Fig. \ref{fig:energy} (c). We use the energy term $E_{BND}$ to tackle this problem.
To compute the boundary term $E_{BND}$, during the rendering of teeth model we change the color of corresponding BOI to green (see Fig.~\ref{fig:energy}d, e). This ensures that when the 2D projection of the roof region is perfectly aligned with the teeth segmentation, the number of pixels with light green color is increased while the number of pixels with dark green color is minimal. Thus, we compute the energy term $E_{BND}$ as:
\begin{equation}
E_{BND} = \#Pixel\_DGreen,
\end{equation}
where $\#Pixel\_DGreen$ refers to the number of dark green pixels in the screen buffer. Please note that the energy term $E_{SP}$ will prevent
the 2D projection of the roof region being pulled into the segmentation too much, since it will increase the number
of white pixels.

To minimize $E$, we combine a gradient descent-like approach with dense sampling which, in practice, ensures good local optima while being computationally efficient. During the descent we estimate the direction of the energy gradient relative to $T$ using finite differences. Along this search direction, we then use golden-section search \cite{Kiefer1953} to find the new optimal value for $T$. To reduce the number of degrees of freedom we first fix translation ($t_x$, $t_y$, $t_z$) and update rotation ($\theta_x$, $\theta_y$, $\theta_z$); then, rotation is fixed and translation updated. We repeat this process until convergence. To avoid getting stuck in poor local minima, we subsequently refine the pose $T$ by sampling the space of possible configurations more densely in a small neighbourhood around the current pose $T$.

To further improve the optimization process we use the pose from previous frame as the initial pose $T$. For the first frame, we bootstrap the process by manually selecting four vertices on the teeth model and find their corresponding positions in two captured images. Through triangulation in stereo analysis, we obtain the corresponding 3D positions for selected vertices; then, an initial guess $T$ for the first frame can be obtained by fitting the initial positions of the selected vertices  to their new positions \cite{Sorkine2009}.


\section{Experimental Results}
\label{sec:result}
The algorithms were implemented in Matlab/C++ and were run on a 2.9GHz Intel Core(TM)i7 processor (32GB RAM) with a single CPU thread.

The teeth performance capture is achieved using three GoPro Hero 5 cameras at $1920\times1080$ resolution, and $24$p linear video mode settings. The performer places his mouth approximately 20 cms from each of the three cameras to allow for optimum coverage and larger overlap between the cameras. Using planar checkerboard patterns, the cameras are calibrated for both intrinsic and extrinsic parameters. The three cameras are synchronized by turning on and off lights. We locate the frames with sudden intensity changes to match the frames from different cameras.
An accurate hardware synchronization could also be employed to eliminate this step.



\noindent
\textbf{Teeth motion tracking.}
On average, the total optimization time is about 15 seconds per frame and the main bottleneck is the computational time for computing the energy function $E$ in Eqn. (\ref{eqn:energy}) which involves counting the number of pixels of specific color in the screen buffer. In our current implementation, we use $940\times480$ screen buffer with 24 bits per pixel, where the pixels are transferred from the GPU into CPU and counted. The computational efficiency can be improved by considering parallel counting on the GPU, e.g., using histogram operation available in CUDA SDK.

\begin{figure}[tb]
\centering
   \includegraphics[width=1.0\linewidth]{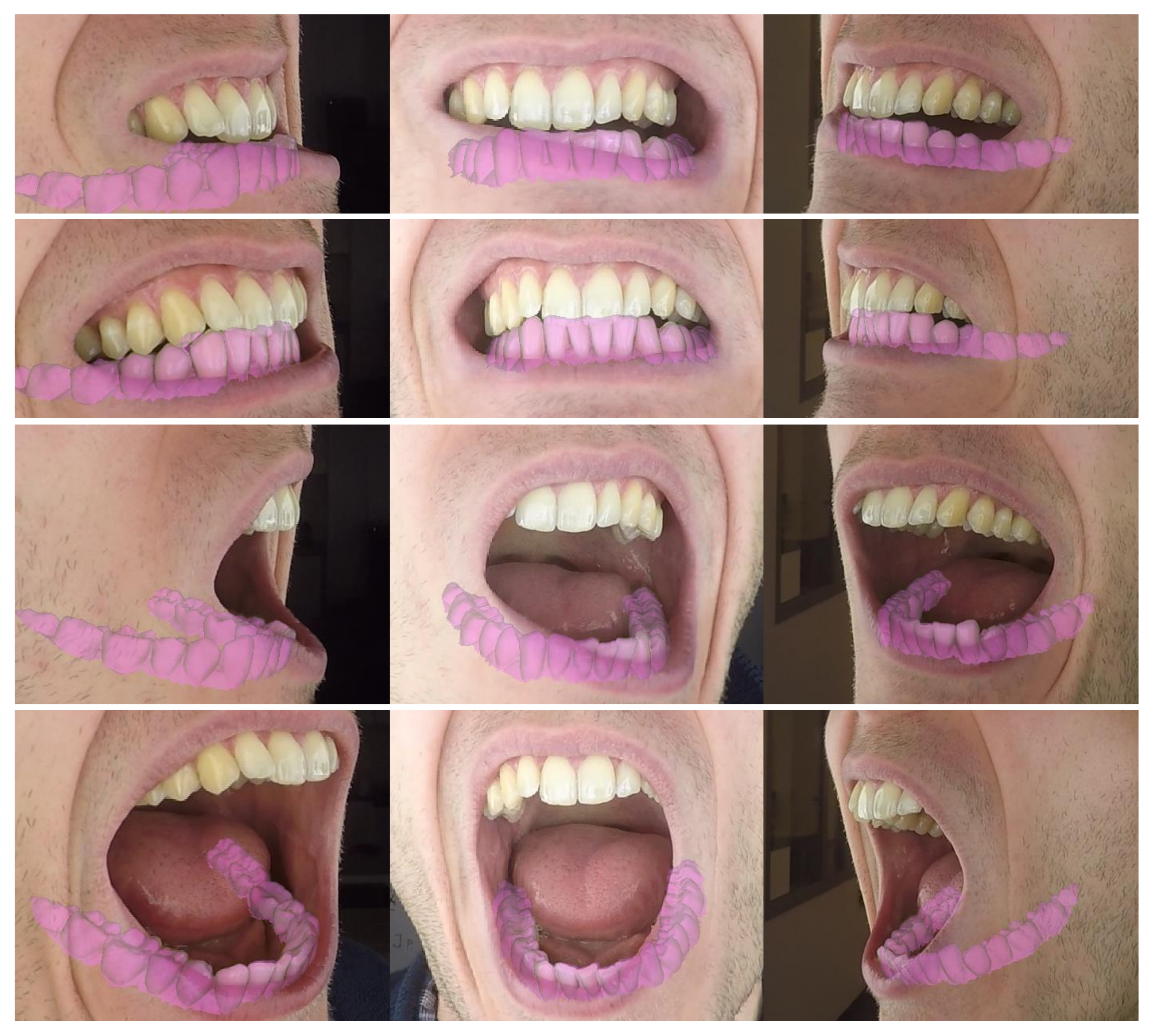}
   \caption{\label{fig:track_result}
   Tracking results from several teeth poses. Each row corresponds to a specific teeth pose and its 2D projections in the three cameras are shown. Note that the 2D projection (pink overlaid region) from the 3D model covers both the visible and the occluded teeth.}
\end{figure}

We recorded $833$ teeth poses of the actor and performed teeth tracking on all these frames for jaw kinematic modeling. We show the results in the supplementary video; a sample of teeth tracking results is shown in Fig.~\ref{fig:track_result}.
\begin{figure}[tb]
\centering
   \includegraphics[width=1.0\linewidth]{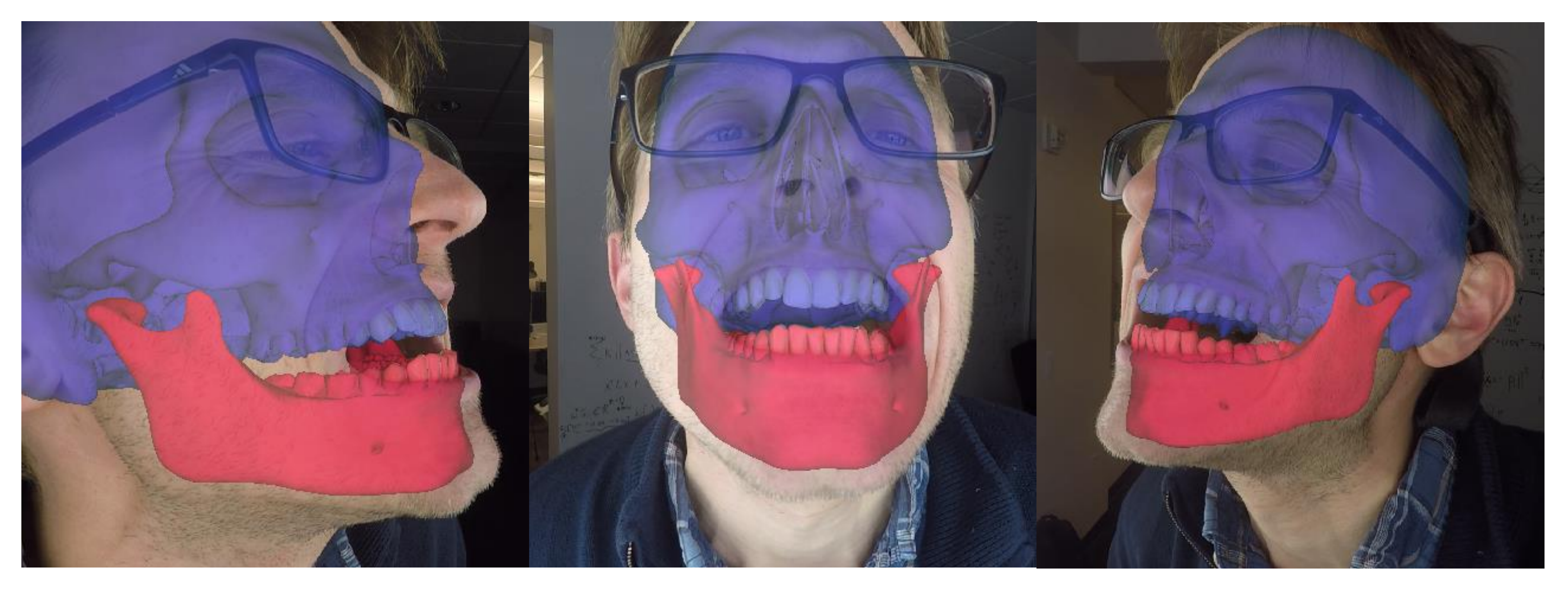}
   \caption{\label{fig:skull}
     The tracked poses of the upper and lower teeth can also be illustratively shown using the projection of the skull (blue) and mandible models (red) on the captured images.}
\end{figure}

\begin{figure}[tb]
\centering
   \includegraphics[width=1.0\linewidth]{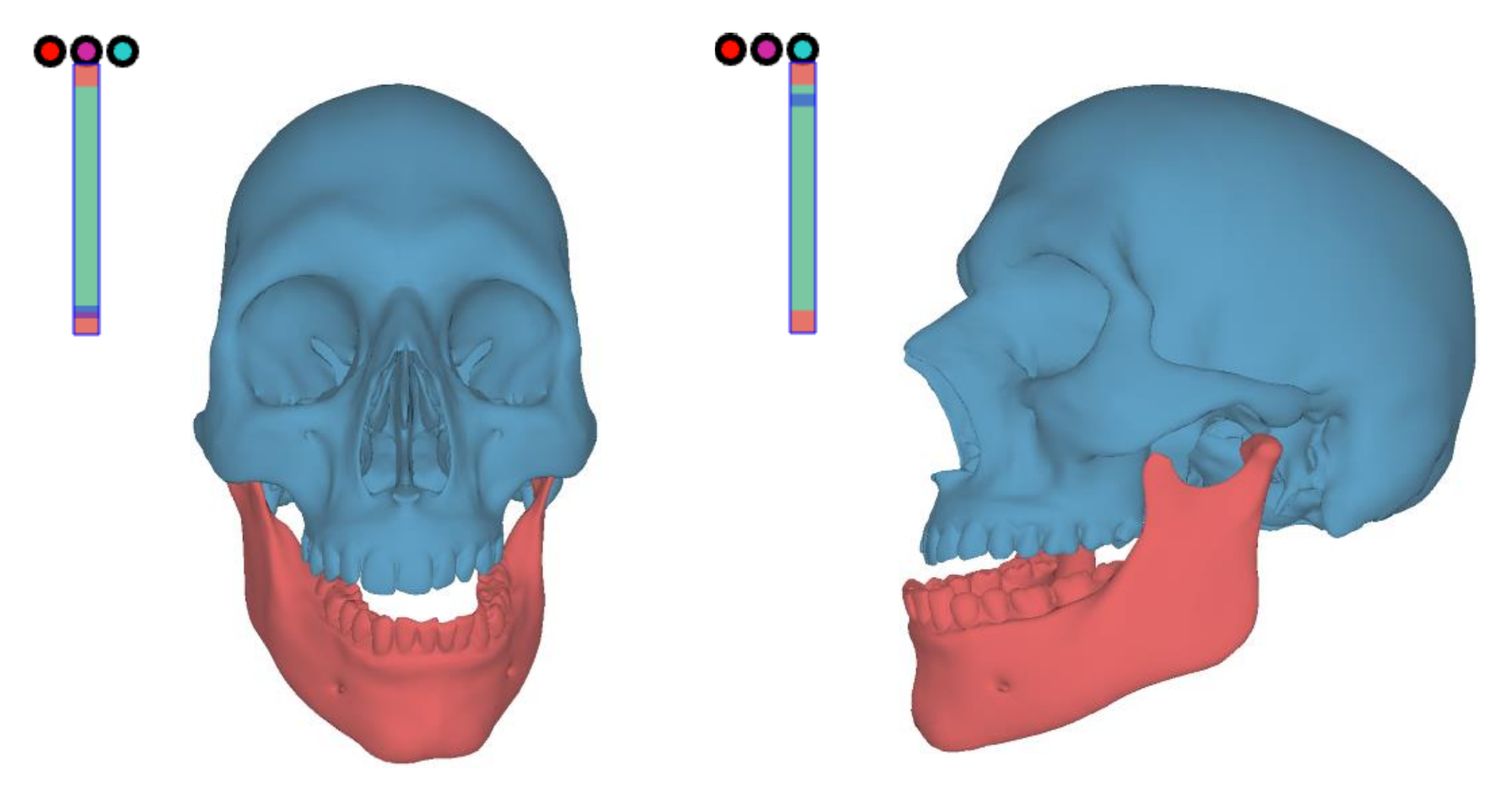}
   \caption{\label{fig:slider}
     Our user interface includes three circle points which correspond to the modes of mouth opening/closing, lateral excursions, and pro/re-trusion, respectively.  When a circle point is selected, a slider appears and the user can move the slider to adjust the corresponding parameter for the desired jaw pose.}
\end{figure}

Error in teeth tracking can come from several sources: calibration, errors in 3D scans of the teeth, and segmentation errors. Unfortunately, it is difficult to get the ground truth for this experiment and therefore, the validation is mainly done qualitatively.
Since we have images from three camera views, we are able to visualize the tracked poses without depth ambiguity, as shown in Fig.~\ref{fig:track_result}. Furthermore, we apply the tracked poses of the upper and lower teeth to the skull and mandible models of the performer, respectively. As shown in Fig.~\ref{fig:skull}, we can create augmented reality sequence by overlaying the skull and mandible models on the captured images. From the reconstructed motions of the skull and mandible models, we can also stabilize skull motions and visualize only the relative motion of the mandible as shown, e.g., in Fig.~\ref{fig:slider}. Note that the skull and mandible models are segmented and reconstructed from the performer's MRI data. While these models are helpful for the visual evaluation of the tracking results, they are not necessary for our tracking approach. It can be seen from the present examples and the supplementary video that the tracked poses of the teeth model are visually well aligned with the real teeth of the performer.


In addition to visual comparison, we provide an approximate metric for measuring the accuracy. Since feature points and their correspondences are difficult to obtain in the case of teeth images, we instead measure the difference in the area from the projection of the teeth model and the teeth in the images. The difference in area can be approximately computed as the ratio between energy term $E_{SP}$ in Eqn. (\ref{eqn:energy_sp}) and the total pixel number of the teeth segmentation. The average error on the captured images is about $0.754\%$.

\textbf{Jaw kinematics model.} With the tracked poses of the performer's lower and upper teeth, the jaw motion extraction, including the data mapping from $SE(3)$ to $se(3)$, can be computed efficiently at the rate of $4.6$ seconds for all the $833$ teeth poses. In Matlab, the nonlinear PCA takes $108$ seconds for learning the jaw kinematics model.

\begin{figure}[tb]
\centering
   \includegraphics[width=1.0\linewidth]{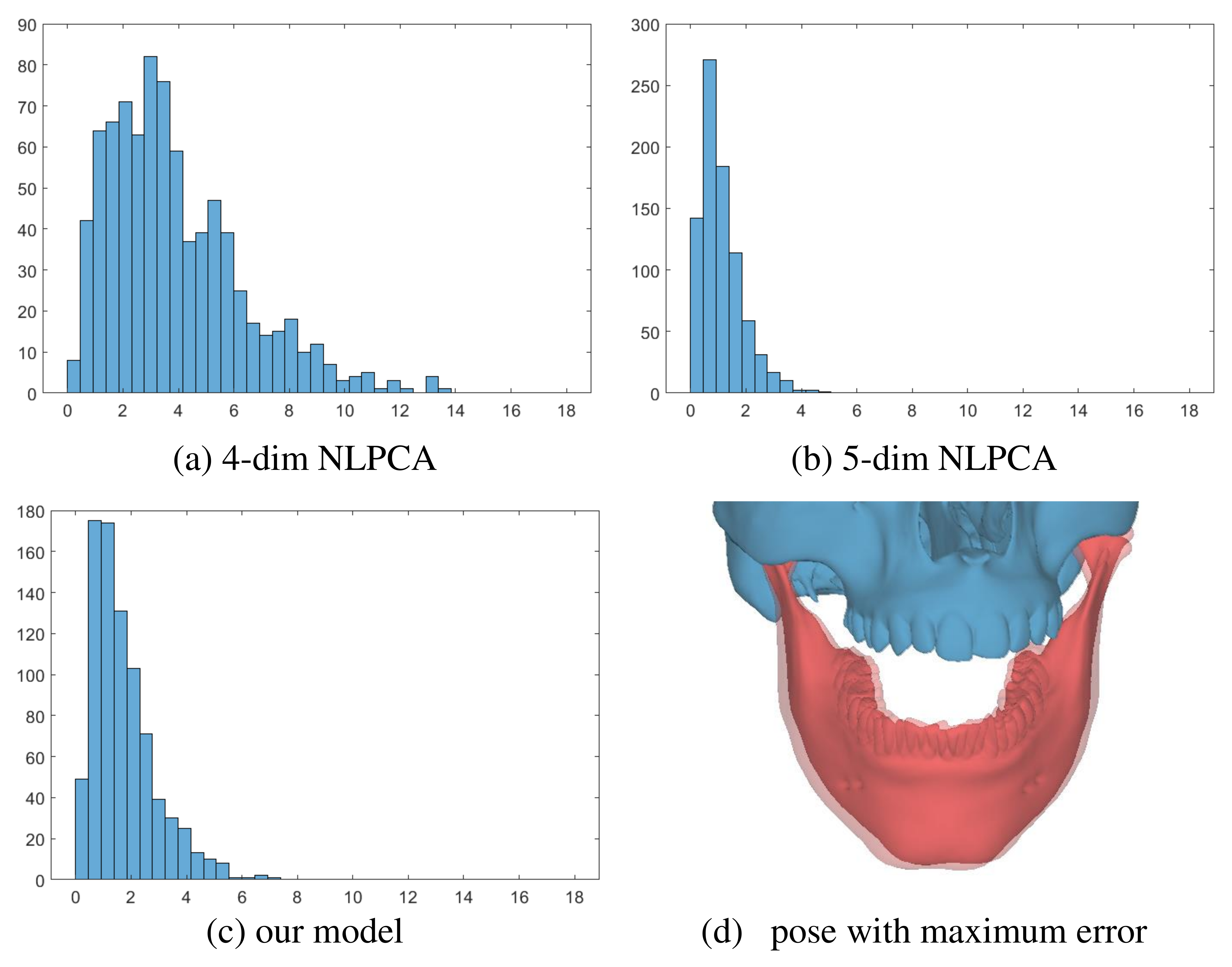}
   \caption{\label{fig:fit_error}
    For each of captured poses, we use NLPCA with different number of component dimensions to reconstruct its original pose in (a) and (b). In (c), the original pose is reconstructed using our final kinematic model according to Eqn.~(\ref{eqn:synthesize}). The histogram of the reconstruction errors of the $833$ captured poses is shown (horizontal axis: error in millimeter; vertical axis: frequency). For the pose with maximum reconstruction error (i.e., 7.2216) using our final kinematics model, its original and reconstructed poses are shown in an overlapping way (d), where the transparent part corresponds to the difference between the two poses.}
\end{figure}

\begin{figure}[tb]
\centering
   \includegraphics[width=1.0\linewidth]{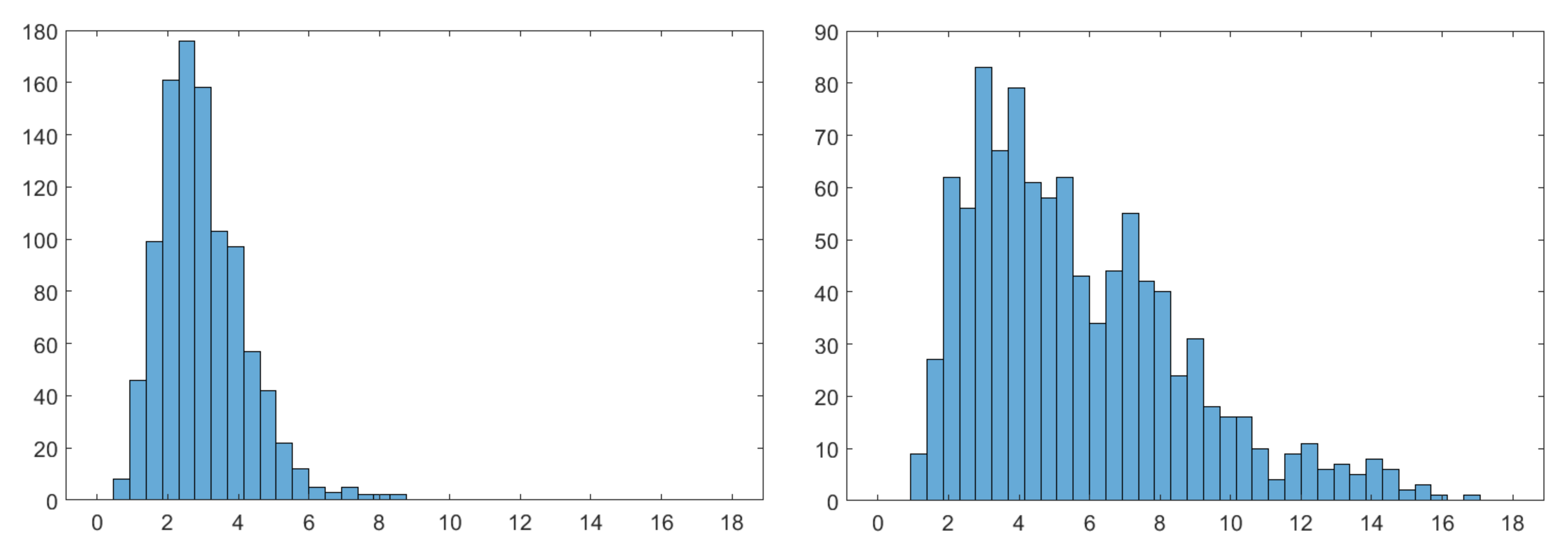}
   \caption{\label{fig:syn_error}
     One thousand jaw poses are randomly synthesized using our kinematics model with and without the dynamic adjustment of the valid ranges for the parameters $d_2$ and $d_3$. The histogram of the accuracy of the $1000$ synthesized poses is given (left is with the dynamic adjustment while right is without the dynamic adjustment), whose layout is similar to that of Fig. \ref{fig:fit_error}.   }
\end{figure}

In Fig. \ref{fig:fit_error}, we use the learned jaw kinematics model to reconstruct the originally captured jaw poses. Then, we measure the reconstruction error by computing the difference between each captured pose and its reconstruction. To measure the difference between a pair of given poses,  we apply them  to the mandible mesh, respectively, and then compute the maximum vertex distance between the two transformed mandible meshes.
As shown in Fig. \ref{fig:fit_error}, our jaw kinematics model can reconstruct the jaw kinematics of a real person up to an error of $1.73$ millimeters on average. To further validate that our jaw kinematics model generates the anatomically realistic poses, we randomly synthesized $1000$ jaw poses using our model. For each synthesized pose, we measure how realistic it is by computing the difference between the pose and its closest one in the real jaw poses that are captured from the performer. The result is shown in Fig. \ref{fig:syn_error}.

Finally, using the jaw kinematics model, we can then synthesize anatomically realistic and visually pleasing jaw poses, i.e., by adjusting the parameters $\{d_i\}_{i=1}^3$ of the model in their respective valid ranges. To facilitate this step, we designed a simple user interface where the user can easily explore the constrained nonlinear space of the jaw kinematics through three sliders, each of which is used to tune a parameter and thus corresponds to a semantically meaningful mode of the jaw motion. An example of this user interface is visible in Fig.~\ref{fig:slider}.

\textbf{Inverse Kinematics (IK).} In addition, we allow the user to generate the jaw poses via direct manipulation. To edit the jaw pose, the user can pick a  vertex on the mandible mesh and mouse-drag it to a new position. The system then automatically updates the parameters of our jaw kinematics model to find the optimal pose that is consistent with this user-specified positional constraint. Given the new position of the selected vertex ($p'$), our goal is to find parameters $\textbf{d}=(d_1, d_2, d_3)$ so that the transformed position of the selected mesh vertex ($p_{\textbf{d}}$) is as close as possible to $p'$:
\begin{equation}
\label{eqn:inverse_jaw}
\arg\min_{\textbf{d}} \lVert p'-p_{\textbf{d}}) \rVert^2 \quad\textrm{subject to valid}\quad \textbf{d}.
\end{equation}
We can efficiently solve Eqn.~(\ref{eqn:inverse_jaw}) with a brute-force approach. We sample the space of possible configurations densely in a small neighbourhood around the current pose and evaluate the best candidate. This takes only few milliseconds. An example of an interactive IK session is presented in the accompanying video.

\begin{figure*}[tb]
\centering
   \includegraphics[width=0.9\linewidth]{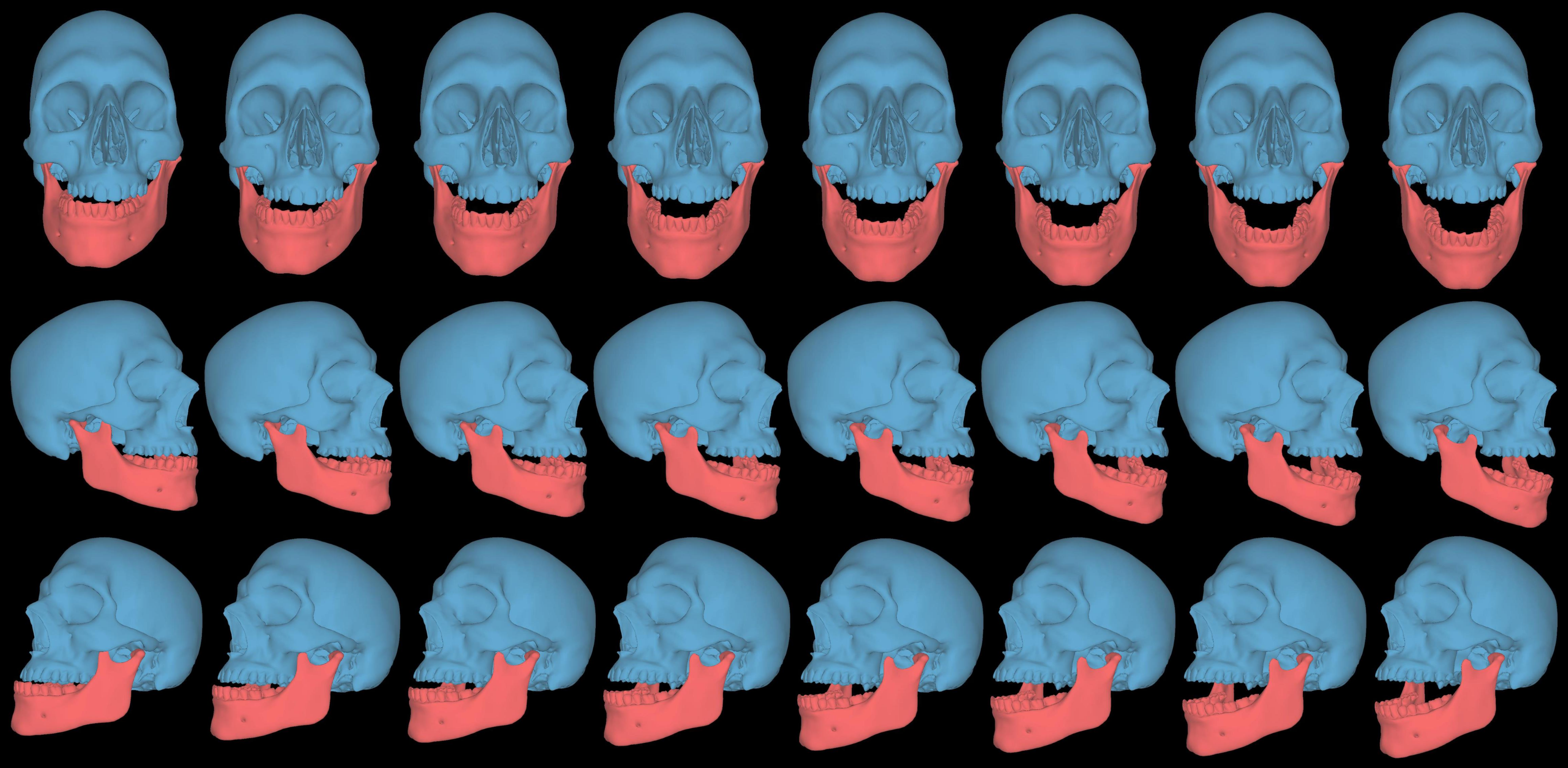}
   \caption{\label{fig:animation}
     Using our interface, the user can create two jaw poses (left and right) and the system automatically generates a jaw animation by keyframe interpolation, where each row corresponds to a different perspective.}
\end{figure*}

\textbf{Animation.} To generate continuous jaw motions, the user can design several key poses of the jaw through our user interface, and then the intermediate poses can be computed using simple linear interpolation between the parameters of the key poses. As shown in Fig.~\ref{fig:animation}, such a simple interpolation scheme already generates high-quality jaw motions which are comparable to real jaw motions.
Furthermore, note that our anatomically realistic jaw kinematics model can be easily integrated into various applications such as anatomically-constrained monocular face capture \cite{wu2016anatomically} or physics-based facial animation \cite{Ichim2017}.

\begin{figure}
    \centering
    \subfigure[]{
    \includegraphics[width=0.20\textwidth]{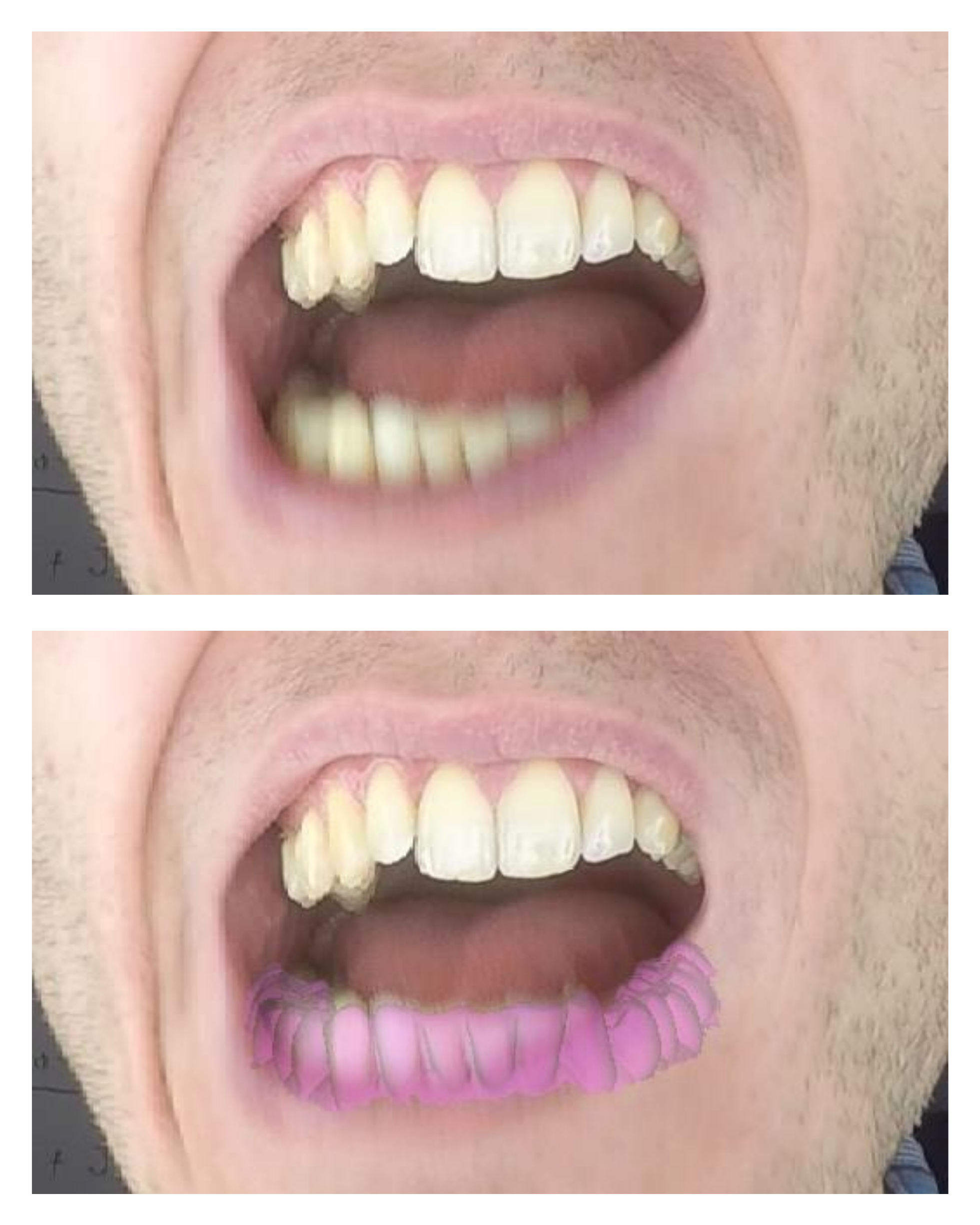}
    \label{fig:fail_1}
    }
    \subfigure[]{
    \includegraphics[width=0.18\textwidth]{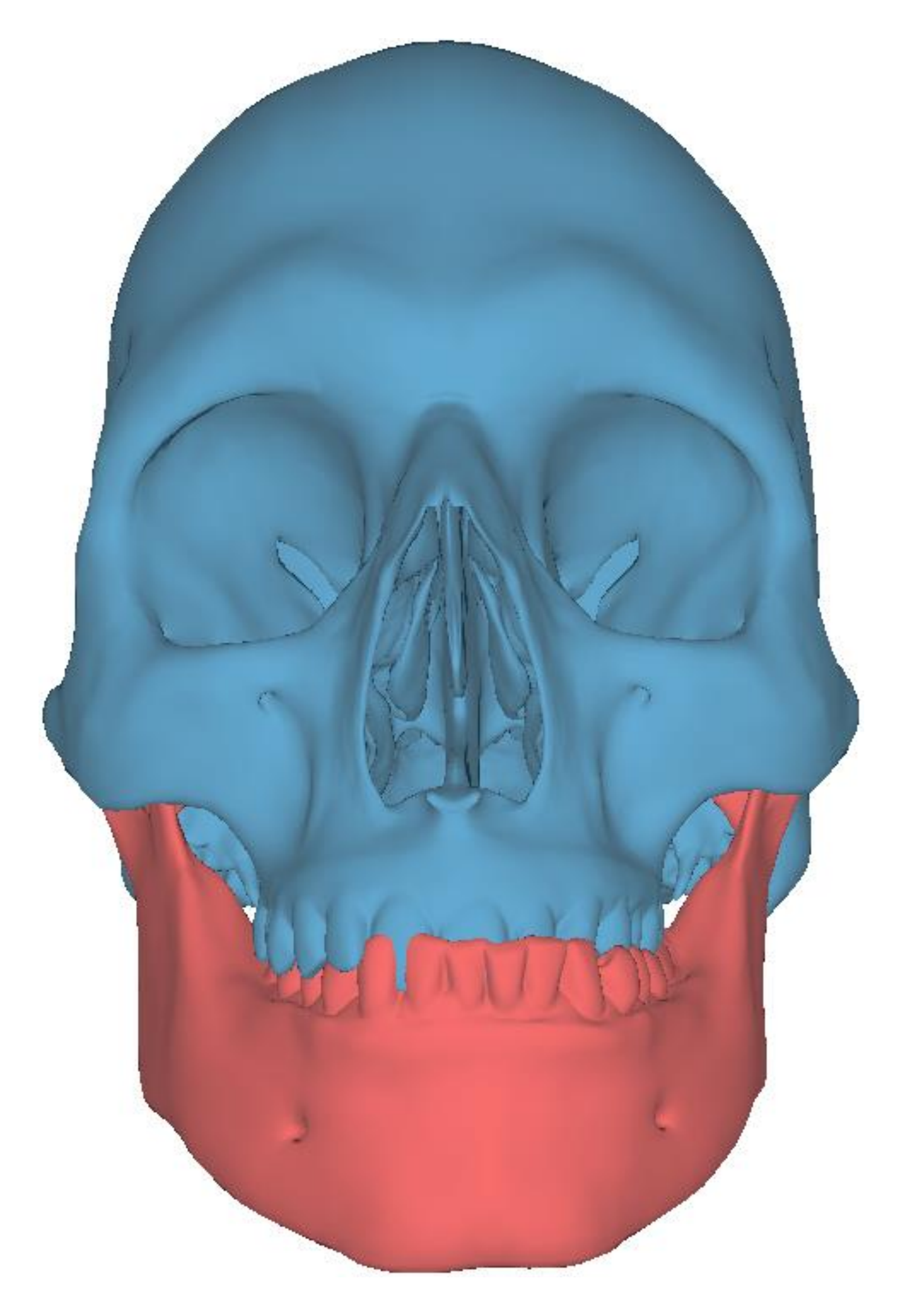}
    \label{fig:fail_2}
    }
    \caption{Limitations and failure cases: (a) We show poor teeth tracking results in the case of images with blur. (b) Invalid jaw poses can be produced when the parameters $\{d_1,d_2,d_3\}$ are chosen near their respective boundary values. In particular, the boundary conditions for parameters $d_2$ and $d_3$ vary and they depend on the current value of $d_1$, However, such boundary conditions are just linear approximations, and may not be accurate.}
    \label{fig:fail}
\end{figure}

\textbf{Limitations.} Rapid motion of the performer can lead to blurred images, which in turn can lead to segmentation errors and incorrect teeth tracking as shown in Fig.~\ref{fig:fail_1}. The use of cameras with higher frame rate can eliminate this problem.

Another limitation stems from the valid ranges of the parameters in our jaw kinematics model. Currently, the ranges of the parameters correspond to a roughly linear approximation of the boundary of the original data points. While such a scheme is simple and practical, invalid jaw poses may still happen when extreme poses are synthesized using the parameters that locate near the boundary of their valid ranges. An example of such scenario is shown in Fig.~\ref{fig:fail_2}, where the mouth is nearly closed, while the jaw is in the rightmost of the lateral excursion and in the maximum of protrusion. A possible solution might be a smooth and tighter bounding volume, e.g., the isosurface representation, as done for the kinematic modeling of joints \cite{Herda2002}.

\section{Conclusion}
In this paper, we build an anatomically realistic jaw kinematics model from data. We use three tripod-mounted GoPro Hero 5 Black cameras to capture the dynamic teeth performances of an actor, from which $833$ jaw poses are tracked and extracted. Then, the jaw kinematics is learned from these jaw poses using the Non-Linear PCA (NLPCA), which effectively captures the nonlinear characteristics of the jaw's motion. The resulting jaw kinematics model has three parameters, each of which corresponds to an intuitive mode of the jaw's motion, i.e., mouth opening, lateral excursions, and pro/re-trusion. Finally, our jaw kinematics model provides an intuitive interface allowing the animators to explore realistic jaw motions in a user-friendly way. Such model is also useful for various application scenarios to guarantee anatomically correct results, such as anatomically-constrained facial animation \cite{wu2016anatomically} or physics-based face simulation \cite{Ichim2017}.

In future, we plan to create a library of user-tailored jaw-kinematics models since each person has her personalized capacity of jaw motion. We plan to obtain completely automatic segmentation algorithm using graph cuts algorithm, where the parameters of the energy function will be learned from the training data with manually annotated segmentation labels. An interesting line of research would be to actually use the learned jaw kinematics model in the teeth tracking problem and vice versa.

\ifCLASSOPTIONcompsoc
  \section*{Acknowledgments}
\else
  \section*{Acknowledgment}
\fi

This material is based upon work supported by the National Science Foundation under Grant Numbers IIS-1617172 and IIS-1622360. Any opinions, findings, and conclusions or recommendations expressed in this material are those of the author(s) and do not necessarily reflect the views of the National Science Foundation. Wenwu Yang was partially funded by the Science and Technology Agency projects of Zhejiang Province (2016C33171). Daniel S\'{y}kora was funded by the Fulbright Commission in the Czech Republic and by the Technology Agency of the Czech Republic under research program TE01020415 (V3C -- Visual Computing Competence Center. We also gratefully acknowledge the support of Research Center for Informatics (No.~CZ.02.1.01/0.0/0.0/16\_019/0000765), Activision and Mitsubishi Electric Research Labs (MERL) as well as hardware donation from NVIDIA Corporation.


\ifCLASSOPTIONcaptionsoff
  \newpage
\fi



%
\bibliographystyle{IEEEtran}
\bibliography{paper_bib}

%




\end{document}